\documentclass[fleqn,usenatbib]{mnras}

\usepackage{newtxtext,newtxmath}

\usepackage[T1]{fontenc}
\usepackage{ae,aecompl}
\usepackage{tabularx}
\usepackage{threeparttable}

\usepackage{graphicx}	
\usepackage{amsmath}	
\usepackage{amssymb}	
\usepackage{ulem}
\usepackage{pifont}

\newcommand{\mvir}{M_\mathrm{v}}
\newcommand{\rvir}{R_\mathrm{v}}

\newcommand{\vvir}{V_\mathrm{v}}
\newcommand{\vmax}{V_\mathrm{max}}
\newcommand{\vpeak}{V_\mathrm{peak}}

\newcommand{\Msun}{\,\textnormal{M}_\odot}

\newcommand{\mpc}{\mathrm{Mpc}}
\newcommand{\Mpc}{\,\textnormal{Mpc}}

\newcommand{\kpc}{\mathrm{kpc}}

\newcommand{\hmpc}{h^{-1}\,\mathrm{Mpc}}

\newcommand{\kms}{{\rm km} \, {\rm s}^{-1}}
\newcommand{\lcdm}{$\Lambda$CDM}

\title[Phat ELVIS]{Phat ELVIS: The inevitable effect of the Milky Way's disk on its dark matter subhaloes} 

\author[T. Kelley et al.]{Tyler Kelley$^{1}$\thanks{E-mail: tkelley1@uci.edu},
James S. Bullock$^{1}$,
Shea Garrison-Kimmel$^{2}$, 
Michael Boylan-Kolchin$^{3}$, \newauthor
Marcel S. Pawlowski$^{1,4}$\thanks{Hubble Fellow},
Andrew S. Graus$^{3}$
\\
$^{1}$Center for Cosmology, Department of Physics and Astronomy, University of California, Irvine, CA 92697, USA\\
$^{2}$TAPIR, California Institute of Technology, Pasadena, CA 91125, USA\\
$^{3}$Department of Astronomy, The University of Texas at Austin, 2515 Speedway, Stop C1400, Austin, TX 78712, USA\\
$^{4}$Leibniz-Institut f\"ur Astrophysik Potsdam (AIP), An der Sternwarte 16, D-14482 Potsdam, Germany\\ 
}

\date{Accepted 2019 May 7. Received 2019 April 29; in original form 2018 November 21}

\pubyear{2018}

\begin{document}
\label{firstpage}
\pagerange{\pageref{firstpage}--\pageref{lastpage}}
\maketitle

\begin{abstract}
We introduce an extension of the ELVIS project to account for the effects of the Milky Way galaxy on its subhalo population.  Our simulation suite, Phat ELVIS, consists of twelve high-resolution cosmological dark matter-only (DMO) zoom simulations of Milky Way-size $\Lambda$CDM~ haloes ($M_{\rm v} = 0.7-2 \times 10^{12}$ M$_\odot$) along with twelve  re-runs with embedded galaxy potentials grown to match the observed Milky Way disk and bulge today.  The central galaxy potential destroys subhalos on orbits with small pericenters in every halo, regardless of the ratio of galaxy mass to halo mass.  This has several important implications. 1)  Most of the {\tt Disk} runs have no subhaloes larger than $V_{\rm max} = 4.5$ km s$^{-1}$ within $20$ kpc and a significant lack of substructure going back $\sim 8$ Gyr, suggesting that local stream-heating signals from dark substructure will be rare.  
2) The pericenter distributions of Milky Way satellites derived from \textit{Gaia} data are remarkably similar to the pericenter distributions of subhaloes in the {\tt Disk} runs, while the DMO runs drastically over-predict galaxies with pericenters smaller than 20 kpc.  3)  The enhanced destruction produces a tension opposite to that of the classic `missing satellites' problem: in order to account for ultra-faint galaxies known within $30$ kpc of the Galaxy, we must populate haloes with $\vpeak \simeq 7$ km s$^{-1}$ ($M \simeq 3 \times 10^{7} \Msun$ at infall), well below the atomic cooling limit of $\vpeak \simeq 16 ~\kms$ ($M \simeq 5 \times 10^{8} \Msun$ at infall).   4) If such tiny haloes do host ultra-faint dwarfs, this implies the existence of $\sim 1000$ satellite galaxies within 300 kpc of the Milky Way. 
\end{abstract}




\section{Introduction}
\label{sec:intro}

A key prediction of standard \lcdm\ cosmology is that dark matter (DM) haloes form hierarchically. This leads to the prediction that massive DM haloes receive a continuous influx of smaller haloes as they grow. Satellite galaxies have been detected around many galaxies and clusters, including the Milky Way (MW), and these are usually associated with the most massive subhaloes predicted to exist. As \lcdm\ cosmological simulations have progressed to higher resolution, it has become clear that the mass spectrum of substructure rises steadily towards the lowest masses resolved \citep[e.g.][]{Aquarius,VL2,GHALO,ELVIS,Griffen2016}. Testing this fundamental prediction stands as a key goal in modern cosmology. The present paper aims to refine existing predictions by including the inevitable dynamical effect associated with the existence of galaxies at the centers of galaxy-size dark matter haloes.

The `missing satellites' problem \citep{Klypin1999MissingSats,Moore1999} points out a clear mismatch between the relatively small number of observed MW satellites and the thousands of predicted subhaloes above the resolution limit of numerical simulations. This discrepancy can be understood without changing the cosmology by assuming that reionization suppresses star formation in the early Universe \citep{Bullock2000,Somerville2002}.  Such a solution matches satellite abundances once one accounts for observational incompleteness \newline
\citep{Tollerud2008,Hargis2014}.   As usually applied, these solutions suggests that haloes smaller than $\sim 5\times 10^8 \Msun$ ($\vmax < 15 ~\kms$, where $\vmax$ is defined as the maximum circular velocity) should be dark \citep{Thoul96,Okamoto08,Ocvirk16,Fitts17,Graus18b}.  

Detecting tiny, dark subhaloes would provide confirmation of a key prediction of \lcdm\ theory and rule out many of the alternative DM and inflationary models that predict a cut-off in the power spectrum at low masses \citep{KL2000,Bode2001,ZB03,Horiuchi2016,Bozek2016,Bose2016}. Since these haloes are believed to be devoid of baryons, they must be discovered indirectly. Within the Milky Way, one promising method for detecting dark subhaloes is via their dynamical effect on thin stellar streams, such as Palomar-5 and GD-1, which exist within $\sim 20$ kpc of the Galactic center \citep[e.g.][and references therein]{Johnston2002,Koposov2010,Carlberg2012,Ngan2015,Bovy17,Bonaca2018}. With future surveys like LSST on the horizon, the number of detected streams around the MW should increase and hold information on the nature of dark substructure.

A statistical sample of MW-like haloes simulated in \lcdm\ with sufficient resolution is necessary to make predictions for these observations. While several such simulations exist in the literature \citep[e.g.][]{Aquarius,VL2,GHALO,Mao2015,Griffen2016}, the vast majority are dark matter only (DMO).  The use of DMO simulations to make predictions about subhalo properties is problematic because DMO simulations do not include the destructive effects of the central galaxy \citep{D10}.  Hydrodynamic simulations show significant differences in subhalo populations compared to those observed in DMO simulations \citep{Brooks14,Wetzel2016,Zhu2016,Sawala2013,bentbybaryons}.  This is particularly true in the central regions of galaxy haloes, where subhaloes are depleted significantly in hydrodynamic simulations compared to DMO counterparts \citep{GarrisonKimmel2017,Despali17,Graus18a}. 

\citet{GarrisonKimmel2017} used the high-resolution hydrodynamic `Latte' simulations \citep{Wetzel2016} to show explicitly that it is the destructive effects of the central galaxy potential, not feedback, that drives most of the differences in subhalo counts between DMO and full-physics simulations.  Their analysis relied on three cosmological simulations of the same halo: 1) a full FIRE-2 physics simulations, 2) a DMO simulation, and 3) a DMO simulation with an embedded galactic potential grown to match the central galaxy formed in the hydrodynamic simulation.  They showed that most of the subhalo properties seen in the full physics simulation were reproduced in the DMO plus potential runs at a fraction of the CPU cost.

In this work, we expand upon the methods of \citet[][GK17 hereafter]{GarrisonKimmel2017} to make predictions for the dark substructure populations of the Milky Way down to the smallest mass scales of relevance for current dark substructure searches ($\vmax \simeq 4.5 ~\kms$).   Unlike the systems examined in GK17, our central galaxies are designed to match the real Milky Way disk and bulge potential precisely at $z=0$ and are grown with time to conform to observational constraints on galaxy evolution.  Using 12 zoom simulations of Milky Way size haloes, we show that the  existence of the central galaxy reduces subhalo counts to near zero within $\sim 20$ kpc of the halo center, regardless of the host halo mass or formation history. This suppression tends to affect subhaloes with early infall times and small pericenters the most. The changes are non-trivial and will have important implications for many areas that have previously been explored with DMO simulations.  Some of these include the implied stellar-mass vs. halo-mass relation for small galaxies \citep[][]{Graus18b,Jethwa18}, quenching timescales \citep{RW18}, ultra-faint galaxy completeness correction estimates \citep{Kim17}, cold stellar stream heating rates \citep{Ngan2015},  predicted satellite galaxy orbits \citep{Riley2018}, and stellar halo formation \citep{Bullock2005,Cooper2010}.  In order to facilitate science of this kind, we will make our data public upon publication of this paper as part of the ELVIS \citep{ELVIS} project site \footnote{\url{http://localgroup.ps.uci.edu/phat-elvis/}}.

In section \ref{sec:sims}, we discuss the simulations and summarize our method of inserting an embedded potential into the center of the host; section \ref{sec:results} explores subhalo population statistics with and without a forming galaxy and presents trends with radius in subhalo depletion. We discuss further implications of our results in section \ref{sec:imp} and conclude in section \ref{sec:disc}.

\begin{table}
\label{tab:mw-values}
\begin{tabular}{lccc}
	Component & Mass & Scale Radius & Scale Height \\
              &	($10^{10}\Msun$) & (kpc) & (kpc) \\[2pt]
    \hline\hline \\[-7pt]
	Stellar Disk & 4.1 & 2.5 & 0.35 \\
    Gas Disk & 1.9 & 7.0 & 0.08 \\
    Bulge & 0.9 & 0.5 &  --- \\
\end{tabular}
\caption{Parameters for the Milky Way potential components at $z = 0$ used in every {\tt Disk} run. The disk scale radii correspond to exponential disk radii, which we model analytically by summing three \citet[][]{Miyamoto1975} disc potentials following \citet{Smith2015}.  The buldge radius corresponds to the scale radius of a \citet{Hernquist90} potential.  Parameters were taken from \citet{McMillan2017} and \citet{BlandHawthorn2016}.} 
\end{table}

\section{Simulations}
\label{sec:sims}

All of our simulations are cosmological and employ the `zoom-in' technique \citep{Katz1993,Onorbe2014} to achieve high force and mass resolution.  We adopt the cosmology of \citet[$\Omega_{\Lambda} = 0.6879, ~\Omega_{\rm m} = 0.3121, ~h = 0.6751$]{Planck2015}.  Each simulation was performed within a global cosmological box of length $50 \,\hmpc = 74.06~\mpc$.  We chose each high-resolution region to contain a single MW-mass ($\sim10^{12} \Msun$) halo at $z = 0$ that has no neighboring haloes of similar or greater mass within $3 \Mpc$. We focus on twelve such haloes, spanning the range of halo mass estimates of the MW summarized in \citet[][]{BlandHawthorn2016}: $\mvir = 0.7-2\times10^{12}\Msun$. Haloes were selected based only on their virial mass  with no preference on merger history or to the subhalo population.  The high-resolution regions have dark matter particle mass of $m_{\rm dm} = 3 \times 10^4 \Msun$ and a Plummer equivalent force softening length of $37$ pc. This allows us to model and identify subhaloes conservatively down to maximum circular velocity $\vmax > 4.5 ~\kms$, which corresponds to a total bound mass $M \gtrsim 5 \times 10^6 \Msun$.

\begin{table*}
\begin{tabular}{lcccccccccc}
	Simulation   &   $\mvir$    &  $\rvir$  & $\vmax$    & $\vvir$    & $N_\mathrm{sub}$  &     $N_\mathrm{sub}$  & $N_\mathrm{sub}$ & $N_\mathrm{sub}$ & $c_{\rm NFW}$ & $z_{0.5}$\\
                 &   ($10^{12} \Msun$)   &  ($\mathrm{kpc})$  & ($\mathrm{\kms}$)  & ($\mathrm{\kms}$) & $<25\,\mathrm{kpc}$ & $<50\,\mathrm{kpc}$  & $<100\,\mathrm{kpc}$ & $<300\,\mathrm{kpc}$ &  & \\[2pt]
	\hline\hline \\[-5pt]
Hound Dog & 1.95 & 330 & 192 & 160 & 118 & 551 & 1858 & 6212 & 10.02 & 1.14 \\ 
Hound Dog \texttt{Disk} & 1.95 & 330 & 202 & 160 & 12 & 213 & 925 & 4351 & 11.82 & 1.31 \\[5pt] 

Blue Suede & 1.74 & 317 & 196 & 154 & 48 & 304 & 1139 & 4368 & 12.36 & 0.74 \\ 
Blue Suede \texttt{Disk} & 1.76 & 319 & 206 & 155 & 4 & 106 & 678 & 3082 & 14.23 & 0.76 \\[5pt] 

Teddy Bear & 1.57 & 307 & 183 & 149 & 62 & 411 & 1562 & 5138 & 10.43 & 0.99 \\ 
Teddy Bear \texttt{Disk} & 1.58 & 307 & 196 & 149 & 4 & 130 & 817 & 3668 & 11.78 & 1.05 \\[5pt] 

Las Vegas & 1.35 & 292 & 175 & 142 & 65 & 336 & 1237 & 4200 & 11.21 & 0.83 \\ 
Las Vegas \texttt{Disk} & 1.40 & 295 & 189 & 143 & 8 & 104 & 644 & 2992 & 13.48 & 0.86 \\[5pt] 

Jailhouse & 1.17 & 278 & 170 & 135 & 71 & 283 & 965 & 3384 & 11.73 & 1.15 \\ 
Jailhouse \texttt{Disk} & 1.20 & 280 & 188 & 136 & 13 & 104 & 486 & 2555 & 15.58 & 1.21 \\[5pt] 

Suspicious & 1.08 & 271 & 158 & 131 & 60 & 339 & 1156 & 3520 & 9.58 & 0.96 \\ 
Suspicious \texttt{Disk} & 1.10 & 272 & 166 & 132 & 10 & 133 & 666 & 2639 & 11.23 & 0.97 \\[5pt] 

Kentucky & 1.09 & 271 & 183 & 132 & 75 & 298 & 899 & 2791 & 18.03 & 1.78 \\ 
Kentucky \texttt{Disk} & 1.08 & 271 & 202 & 131 & 9 & 85 & 365 & 1761 & 24.15 & 2.22 \\[5pt] 

Lonesome & 1.02 & 265 & 159 & 129 & 91 & 378 & 1154 & 3390 & 11.14 & 1.56 \\ 
Lonesome \texttt{Disk} & 1.04 & 267 & 180 & 130 & 5 & 121 & 494 & 2164 & 16.54 & 1.55 \\[5pt] 

Tender & 0.95 & 259 & 152 & 126 & 74 & 344 & 1070 & 3190 & 10.16 & 0.81 \\ 
Tender \texttt{Disk} & 0.96 & 260 & 171 & 126 & 7 & 97 & 448 & 2112 & 16.05 & 0.84 \\[5pt] 

Hard Headed & 0.85 & 250 & 160 & 121 & 97 & 492 & 1389 & 3296 & 14.54 & 1.79 \\ 
Hard Headed \texttt{Disk} & 0.89 & 253 & 179 & 123 & 14 & 175 & 782 & 2412 & 18.65 & 1.76 \\[5pt] 

Shook Up & 0.72 & 236 & 147 & 115 & 92 & 346 & 1007 & 2740 & 12.16 & 1.46 \\ 
Shook Up \texttt{Disk} & 0.73 & 238 & 173 & 115 & 8 & 138 & 561 & 1767 & 20.67 & 1.51 \\[5pt] 

All Right & 0.65 & 229 & 140 & 111 & 66 & 328 & 898 & 2544 & 12.02 & 1.69 \\ 
All Right \texttt{Disk} & 0.71 & 235 & 164 & 114 & 5 & 116 & 479 & 1765 & 17.10 & 1.28 \\[5pt] 
\end{tabular}
\label{tab:sim-info}
\caption{Discography of halo properties at $z=0$.  Haloes are listed in pairs corresponding to DMO (first) and those run with embedded galactic potentials (second,  designated  `{\tt Disk}').
The remaining
columns list the \citet{Bryan1998} virial mass, virial radius, maximum circular velocity, $\vmax$, 
and virial velocity ($\sqrt{G\mvir/\rvir}$), along with the the total number of subhaloes with
$\vmax > 4.5\kms$ that survive to $z=0$ within $25$, $50$, $100$, and $300~\kpc$ of the
halo center, the concentration based off of a best fit NFW, and the redshift at which the host obtained 50\% of its final mass.}
\end{table*}

We ran all simulations using \texttt{GIZMO} \citep{GIZMO}\footnote{\url{http://www.tapir.caltech.edu/~phopkins/Site/GIZMO.html}}, which uses an updated version of the TREE+PM gravity solver included in \texttt{GADGET-3} \citep{Springel2005}. We generated initial conditions for the simulations at $z = 125$ using \texttt{MUSIC} \citep{MUSIC} with second-order Lagrangian perturbation theory. We identify halo centers and create halo catalogs with \texttt{Rockstar} \citep{rockstar} and build merger trees using \texttt{consistent-trees} \citep{ctrees} based on 152 snapshots spaced evenly in scale factor. The merger trees and catalogs allow us to identify basic halo properties at each snapshot, including the maximum circular velocity $\vmax$ and virial mass $\mvir$ for the main progenitor of each host halo and subhalo.  For each subhalo, we record the time it first fell into the virial radius of its host and also the largest value of $\vmax$ it ever had over its history, $\vpeak$.  In most cases $\vpeak$ occurs just prior to first infall.

For the embedded disk galaxy simulations, we insert the galaxy potentials at $z = 3$ ($t_{\rm lookback} \sim 11.7$ Gyr), when galaxy masses are small compared to the main progenitor (typically, $M_{\rm gal}/\mvir \,(z=3) \simeq 0.03$).  Prior to $z=3$, the {\tt Disk} runs and DMO simulations are identical. At $z=3$, we impose the galaxy potential, which is centered on a sink particle with softening length 0.5 kpc and mass $10^8 \Msun$.  The sink particle is initially placed in the center of the host halo, as determined by \texttt{Rockstar}.  We have found that dynamical friction keeps the sink particle (and thus the galaxy potential) centered on the host halo throughout simulations -- with a maximum deviation from center of  $\sim 150$ pc at $z=0$. Host halo mass accretion rates, positions, and global evolution are almost indistinguishable from the DMO runs after the galaxy potentials are included.    As discussed below, the galaxy potential grows with time in a way that tracks dark matter halo growth.  All galaxy potentials at $z=0$ are the same, with properties that match the Milky Way today, as summarized in Table \ref{tab:mw-values}.  This means that our higher $\mvir$ halos will have smaller $M_{\rm gal}/\mvir$ ratios, where $M_{\rm gal} = M_{\rm stellar \, disk} + M_{\rm gas \, disk} + M_{\rm bulge}$.  The full range of our suite it $M_{\rm gal}/\mvir \simeq 0.035 - 0.1$.

\begingroup
    \setlength{\tabcolsep}{2pt}
    \begin{figure*}
    \begin{tabular}{cc}
	    \includegraphics[width=\columnwidth]{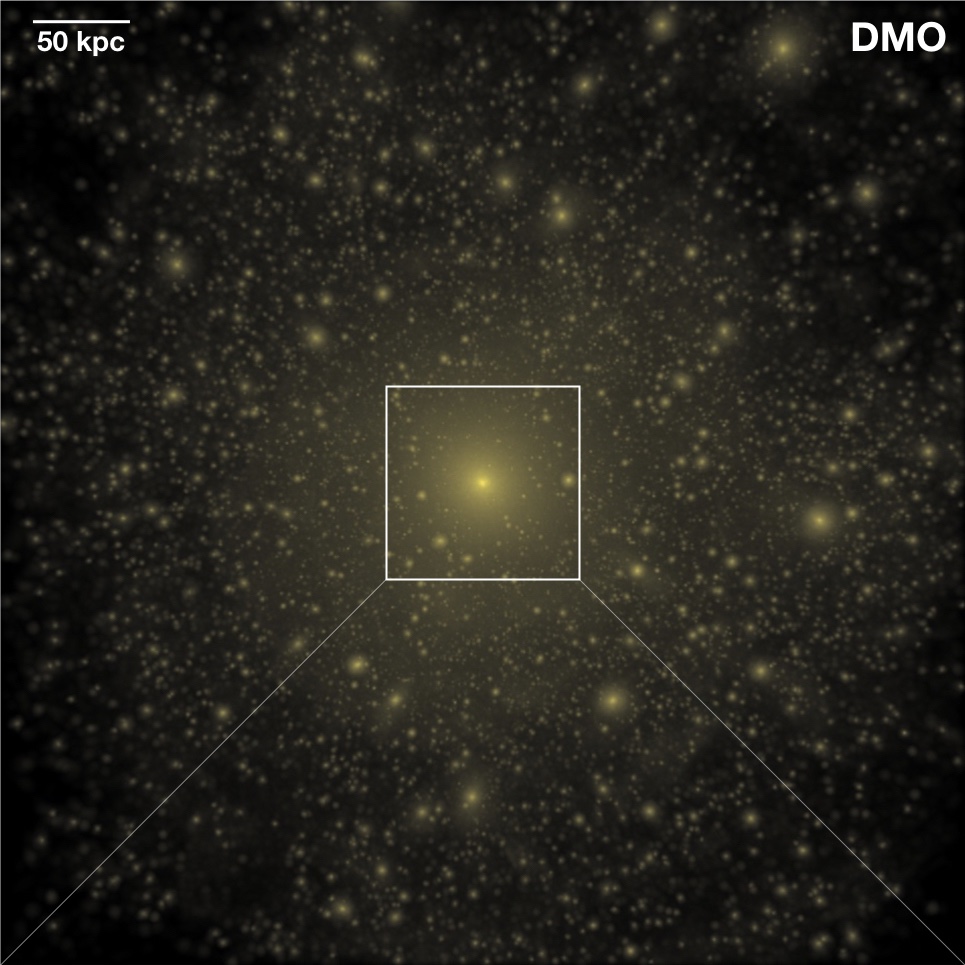} &
        \includegraphics[width=\columnwidth]{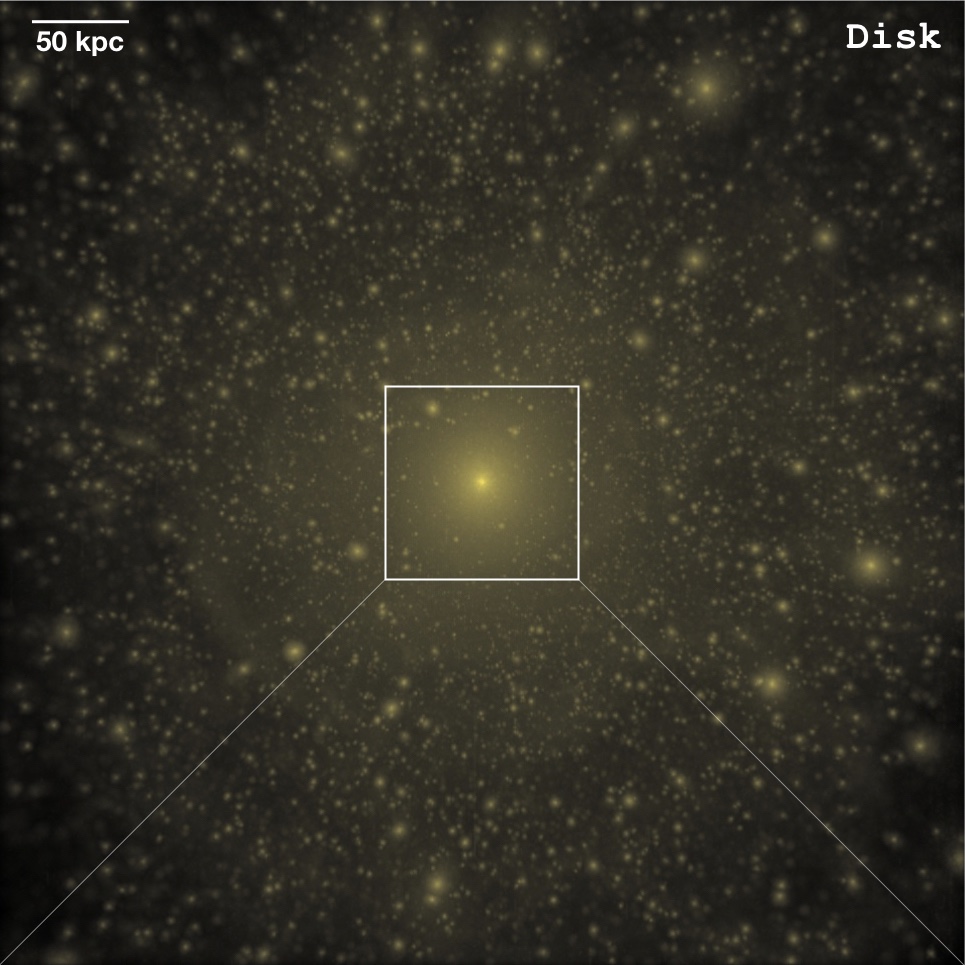} \\
	    \includegraphics[width=\columnwidth]{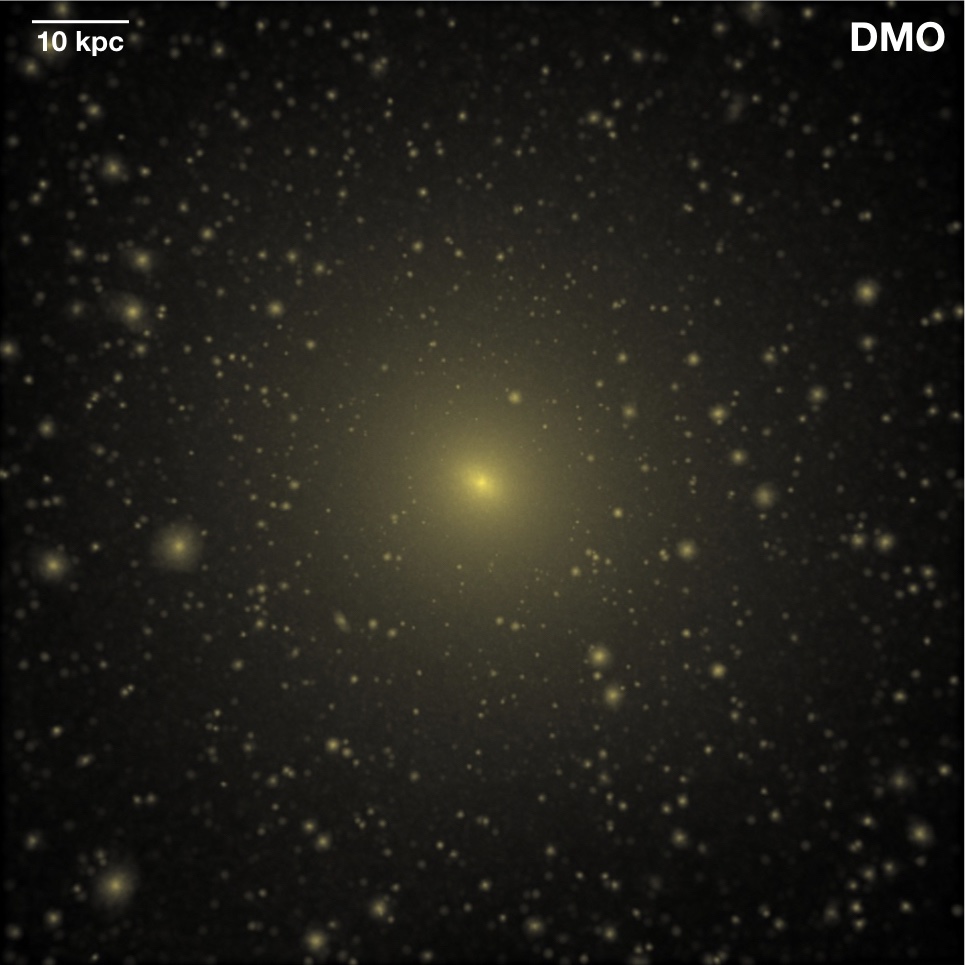} &
	    \includegraphics[width=\columnwidth]{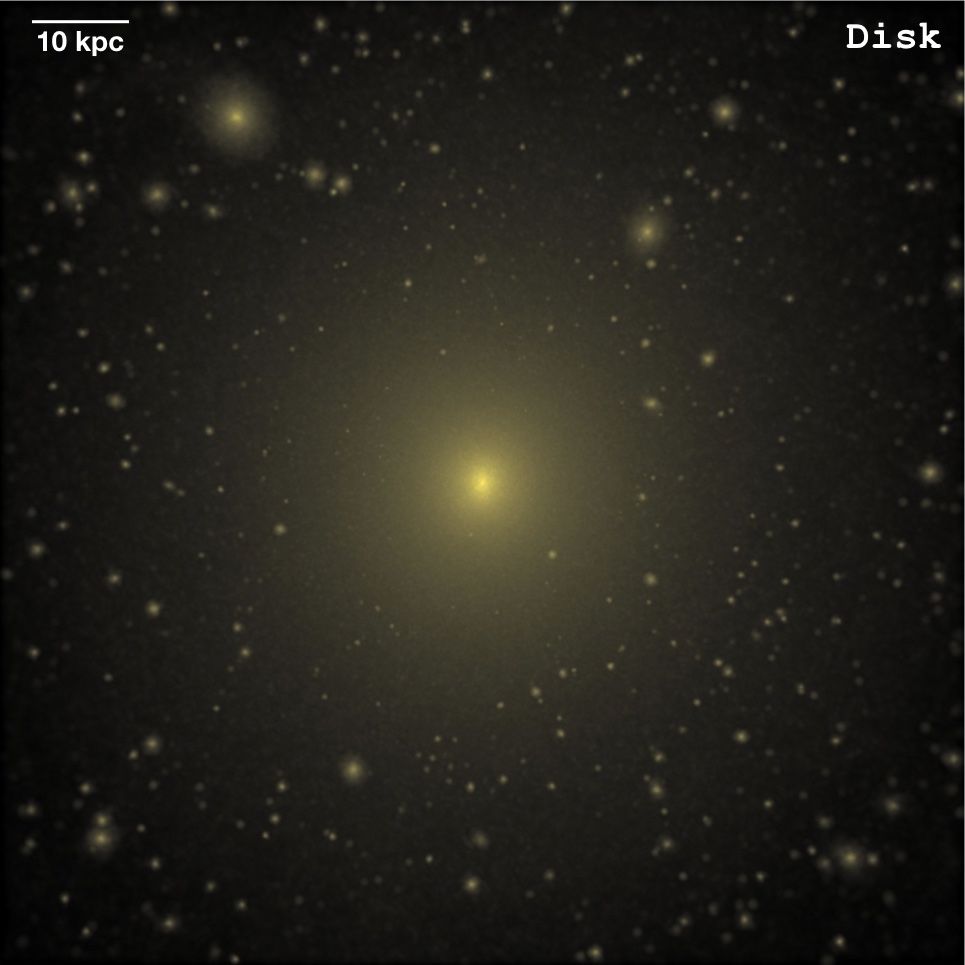} \\
    \end{tabular}
    \caption{Visualization of the dark matter for Kentucky (left) and Kentucky {\tt Disk} (right).  The top panels span 500 kpc, approximately the virial volume of this halo.  The bottom panels span 100 kpc.  The absence of substructure at small radii in the {\tt Disk} runs is striking.  An enhancement in central dark matter density is also seen in the {\tt Disk} runs, which is a result of baryonic contraction. The disc potentials are oriented face-on in these images.}
    \label{fig:density}
    \end{figure*}
\endgroup

The properties of our twelve pairs of host haloes, along with the number of resolved subhaloes
identified by \texttt{Rockstar} within several radial cuts of that host, are listed in
Table \ref{tab:sim-info}. The first column lists the name of each simulated halo.  The names are inspired by the twelve greatest\footnote{As determined scientifically using Bayesian statistics and ideas motivated by string theory.} songs recorded by the Elvis Presely over his 24 year musical career. Haloes are listed in DMO/disk-run pairs, such that the disk simulations are identified with an added `{\tt Disk}' to the name. Virial masses and radii (columns 2 and 3) use the \citet{Bryan1998} definition of virial mass.  Columns 4 and 5 list $\vmax$ and virial velocity, $\vvir$.  Columns 6-9 give the cumulative count of subhaloes with $\vmax > 4.5 ~\kms$ within 25, 50, 100, and 300 kpc of each host's center.  As we discuss below, the difference in subhalo counts between the {\tt Disk} runs and DMO runs is systematic and significant, especially at small radii. Column 10 lists the best-fit \citet[][NFW]{NFW} concentration for each halo.  Note that the {\tt Disk} runs are always more concentrated, even though their formation times (column 11) are similar.  This particularly true of the lower mass host halos.  The reason is that the dark matter in the host haloes contract in response to the central galaxy. 

Throughout this work we characterize subhaloes in terms of their $\vmax$ and $\vpeak$ (peak $\vmax$).  We do this because we have found $\vmax$ selection to produce more consistent results between halo finders (e.g. \texttt{Rockstar} and \texttt{AHF}) than mass selection (for subhaloes in particular, mass definitions are more subjective).  For reference, \citet{ELVIS} found median relations between velocity and mass of 
$M_{\rm peak}/M_{\odot} \simeq 9.8 \times 10^{7} (\vpeak/10 ~\kms)^{3.33}$
and
$M/M_{\odot} \simeq 9.1 \times 10^{7} (\vmax/10 ~\kms)^{3.45}$.

\begin{figure*}
\begin{tabular}[t]{cc}
	 & \\ [-22pt] 
	\includegraphics[width=7.4cm]{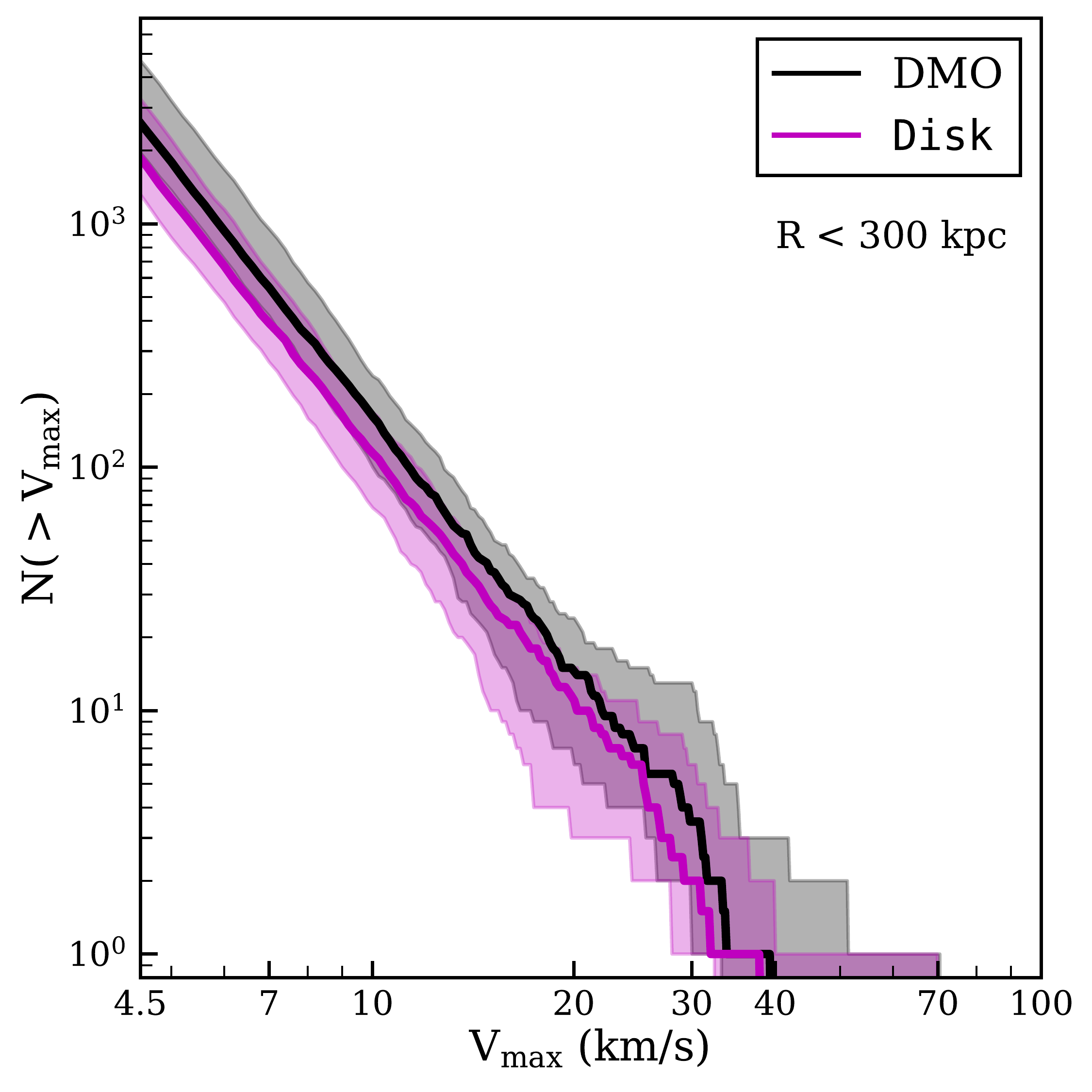} &
    \includegraphics[width=7.4cm]{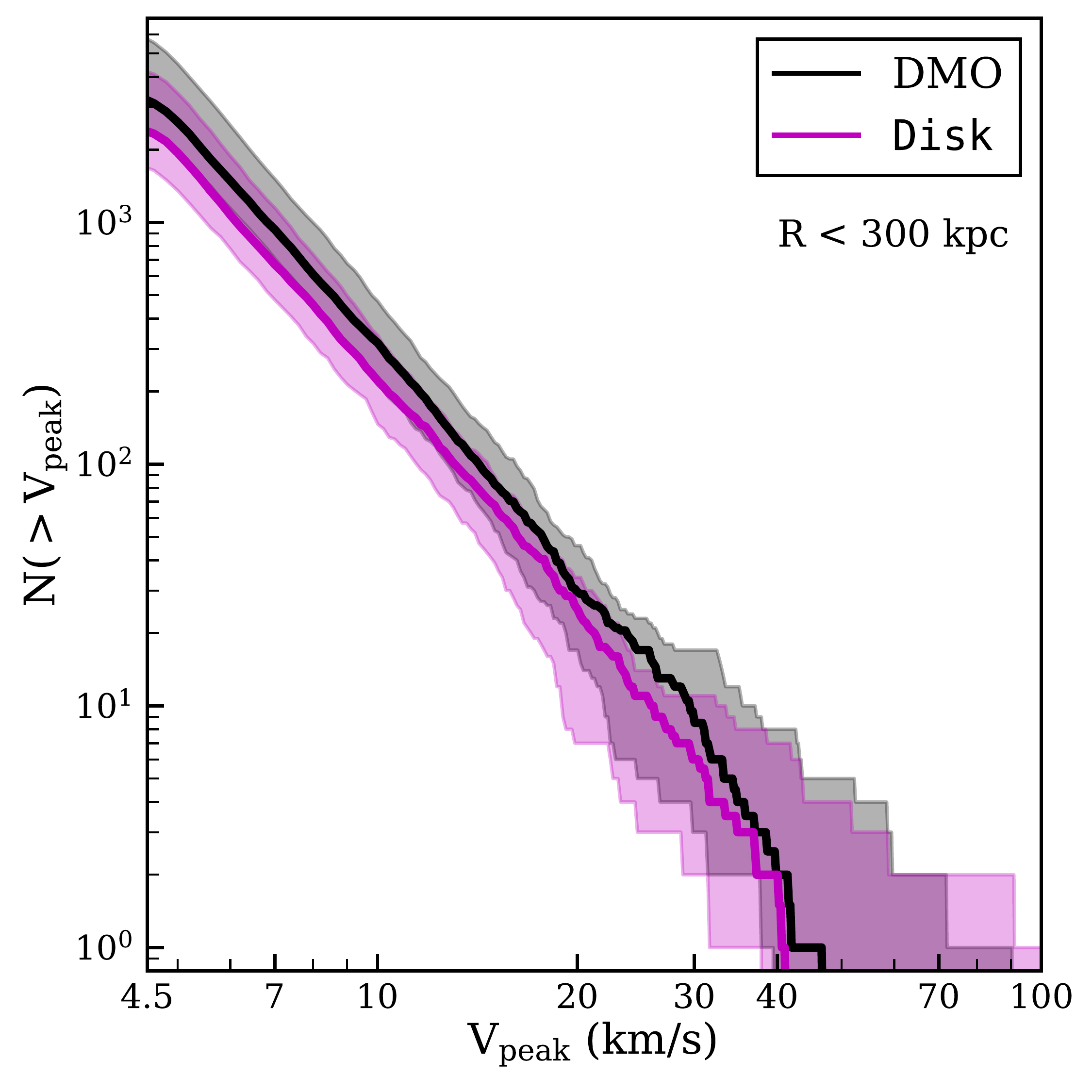} \\ [-7pt]
    \includegraphics[width=7.4cm]{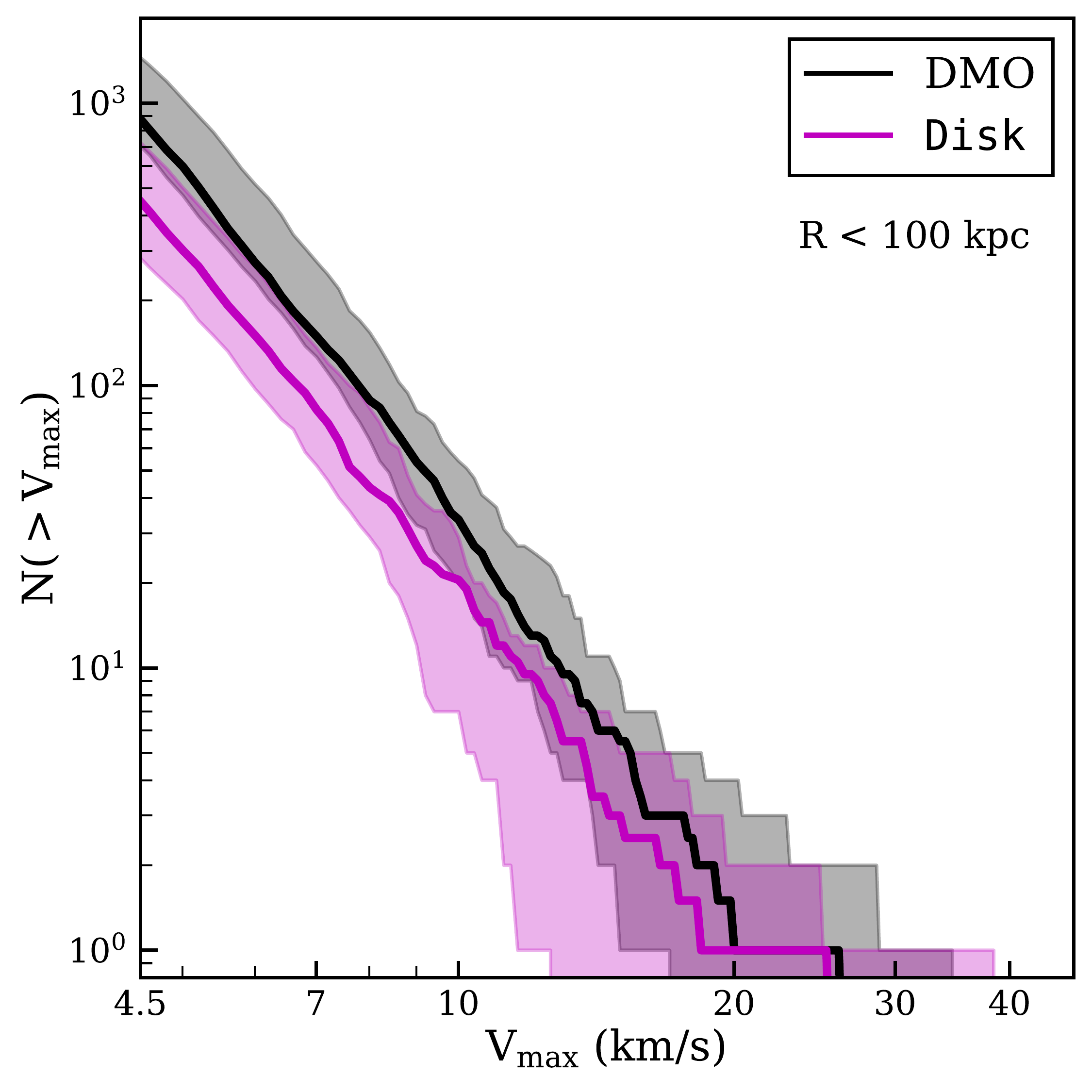} & 
    \includegraphics[width=7.4cm]{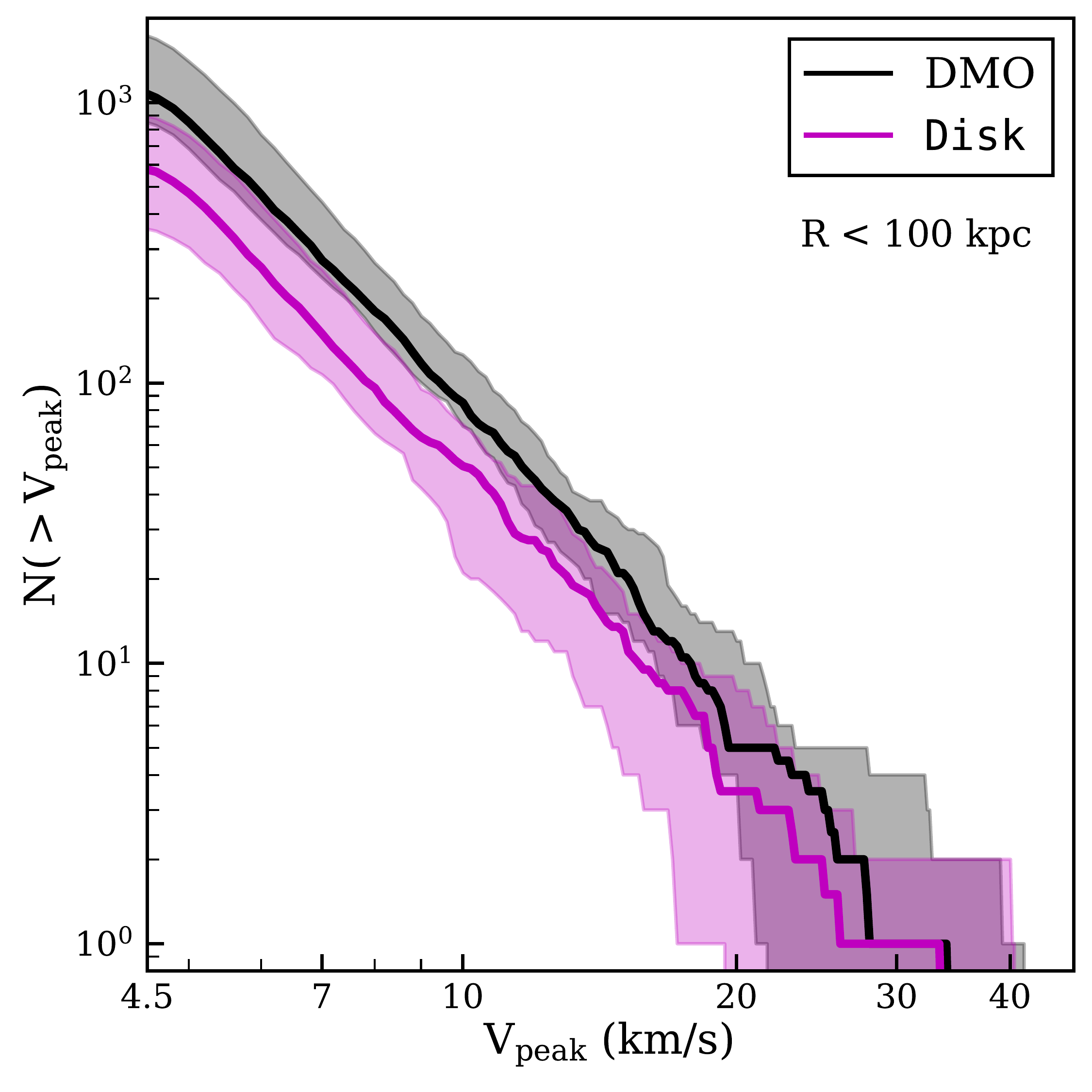} \\ [-7pt]
    \includegraphics[width=7.4cm]{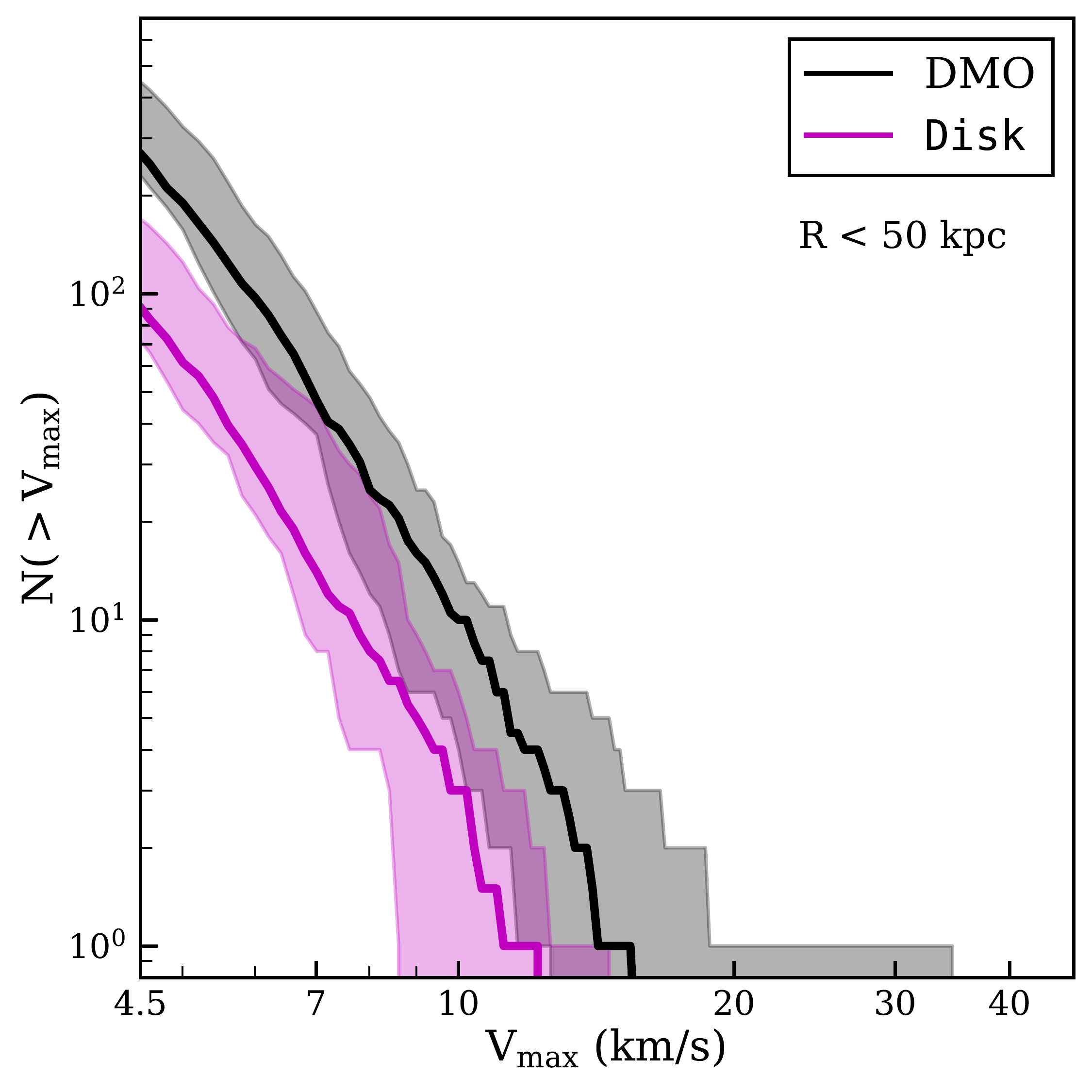} & 
    \includegraphics[width=7.4cm]{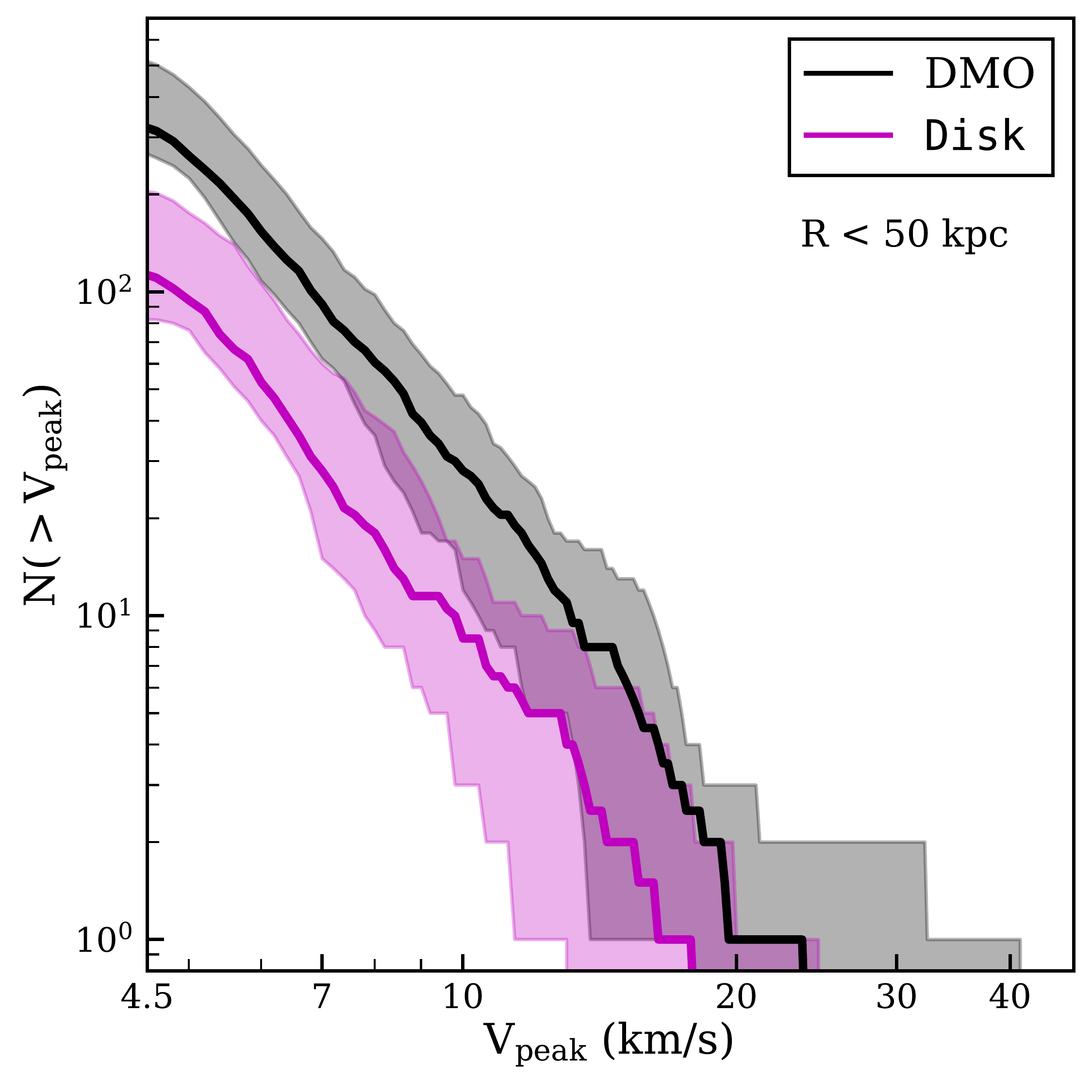} \\
\end{tabular}
    \caption{Cumulative $\rm \vmax$ (left) and $\rm \vpeak$ (right) distributions for subhaloes within $R=300$, $100$, and $50$ kpc (top to bottom) for all 12 of our DMO (black) and DMO+{\tt Disk} runs (magenta).  The solid lines are medians while the shaded bands span the full extent of the distributions.  Note that the roll-off at low $\vpeak$ in the right panels are signatures of incompleteness.  The $\vpeak$ completeness limit gets worse as we approach the halo centres (where stripping is more important). The simulations appear reasonably complete to $\vpeak \simeq 5 \kms$ within $300$ kpc.  This limit drops to $\vpeak \simeq 6 ~\kms$ within $50$ kpc. There is no such roll-off in $\vmax$, which suggests we are complete down to $\vmax \simeq 4.5 ~\kms$ throughout the haloes.}
    \label{fig:vfunc}
\end{figure*}

\begin{figure*}
	\includegraphics[width=\columnwidth]{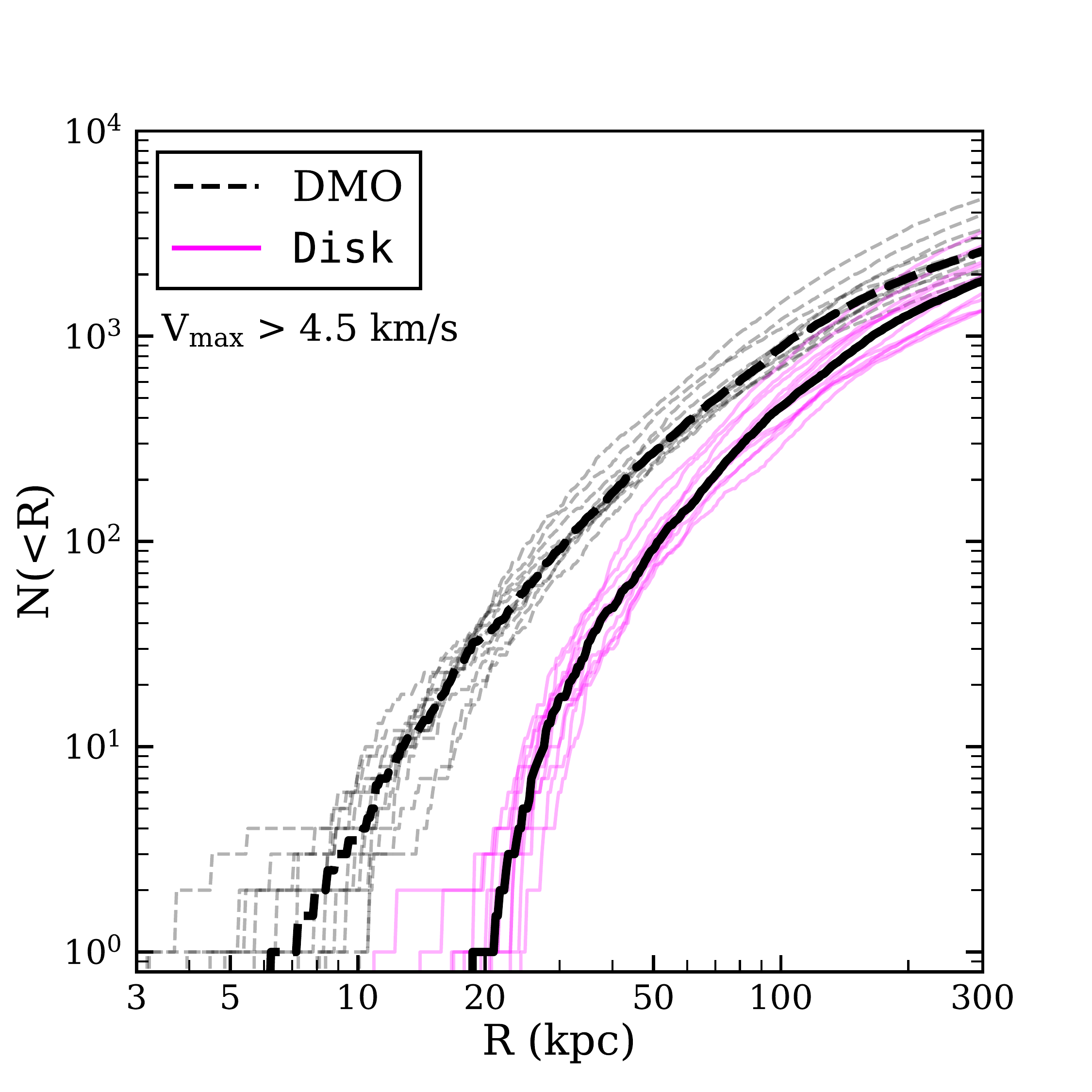}
    \includegraphics[width=\columnwidth]{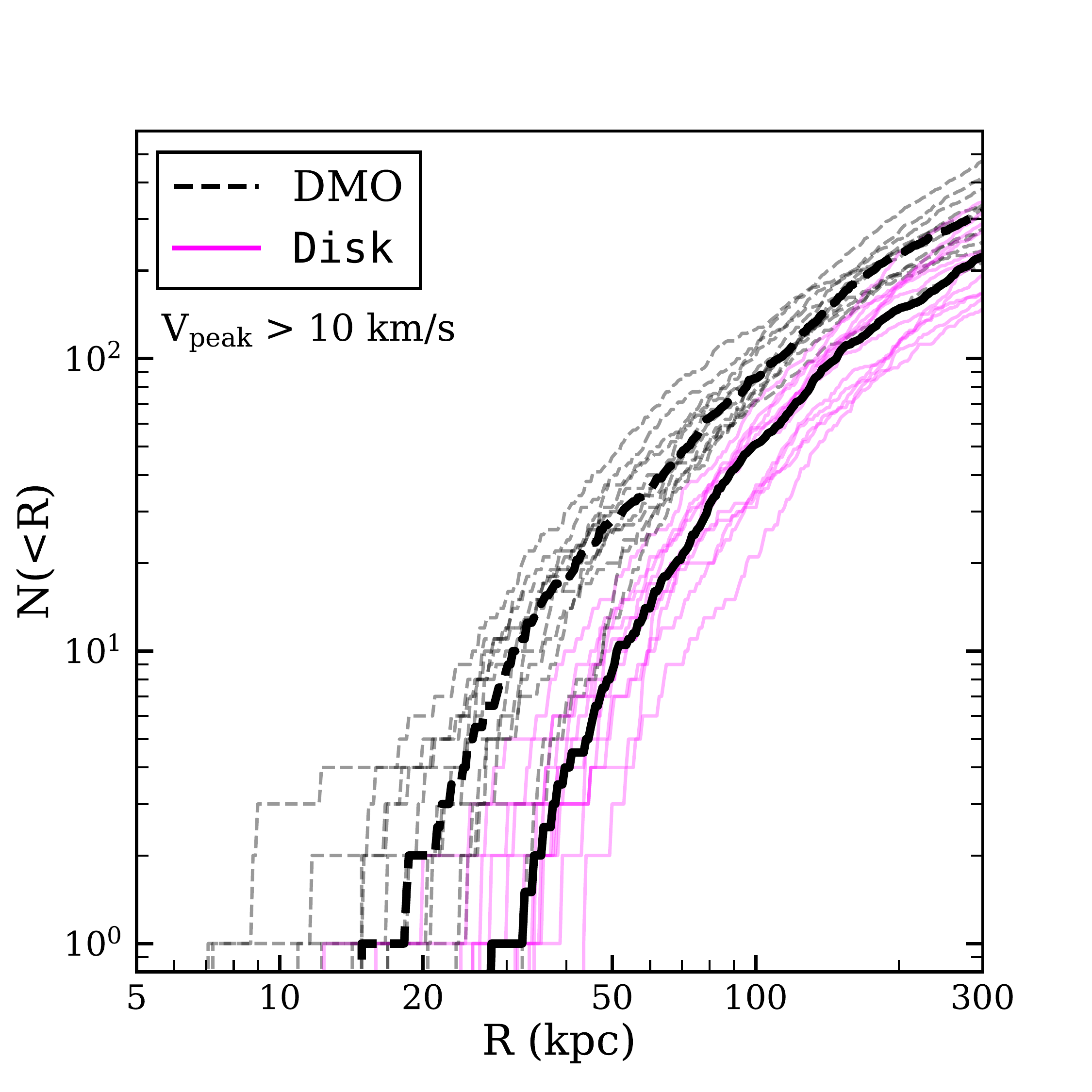}
    \caption{Cumulative counts for subhaloes within a given radius of the host. The black dashed lines represent the dark matter only runs and the magenta lines represent the same haloes with an embedded MW-like potential. Left: subhaloes with $\vmax > 4.5 ~\kms$. Right: subhaloes with $\vpeak > 10 ~\kms$.  Note that the vertical axis scales are significantly different on the left and right.}
    \label{fig:rdist}
\end{figure*}

\subsection{Embedded Potentials}
\label{ssec:potentials}

The effects of the central baryonic disk is included in the DMO simulations following the basic technique described in \citet{GarrisonKimmel2017}. Our embedded potentials are more detailed than those used by GK17 in order to more accurately model the MW galaxy. Specifically, we include an exponential stellar disk, an exponential gaseous disk, and a Hernquist bulge component.  The galaxy potentials evolve from high redshift using empirically-motivated scaling relations (see \ref{sssec:evolution}) and we force them to match currently observed MW properties at $z = 0$. These $z=0$ properties are taken from \citet{McMillan2017} and \citet{BlandHawthorn2016} as summarized in Table \ref{tab:mw-values}.  For simplicity, we hold all disk orientations fixed throughout the simulation. The analysis in section 3.4 of GK17 suggests that the results do not largely depend on the orientation or shape of the embedded potential.

\subsubsection{Modeling Evolution}
\label{sssec:evolution}
We allow the galaxy potential to evolve with time by letting it track the dark matter halo growth using abundance matching \citep[AM,][]{Behroozi2013}.  We enforce a constant offset in stellar mass at fixed halo mass such that the $z=0$ galaxy mass matches the desired MW stellar disk mass at $z=0$ for each of the simulations. Note that each halo has a different $z=0$ virial mass (Table \ref{tab:sim-info}) and this means that each one has a different offset from the mean AM relation throughout its history. If it is low at $z=0$, it is low at $z=3$ and vice versa.  However, while each galaxy/halo has a distinct growth rate, all of them end up the same observationally-constrained `Milky Way' galaxy at $z=0$.

The scale radii at higher redshift are matched to median results from CANDELS, specifically those listed in Table 2 of \citet{vanderWel2014}. The scale height is adjusted to keep the ratio between the scale length and height constant throughout time, with the $z = 0$ ratio as the chosen value. While this will keep the proportions of galaxy components constant, the overall size of the galaxy grows with time as informed by observations. The galaxy mass evolution for one of our hosts is shown in Figure \ref{fig:growgal} for reference. 

\subsubsection{Stellar Disk}
\label{sssec:star-disk}
The stellar disk of most galaxies is well represented with an exponential form \citep{Freeman1970}. However, the potential for such a distribution cannot be derived analytically. An alternative analytic potential commonly used is the \citet[][MN]{Miyamoto1975} disk potential: 
\begin{equation}
\Phi(R,z) = \frac{GM_d}{\sqrt{R^2 + \left(R_d + \sqrt{z^2 + b^2}\right)^2}}
\label{eq:MN}
\end{equation}
where $M_d$ is the total disk mass, $R_d$ is the scale length, $b$ is the scale height, and $R$ and $z$ are the radial and vertical distances from the center, respectively. 

Unfortunately, a single MN disk is a poor match to an exponential disk. The surface density in the center is too low and the surface density too high at large radii. A better approximation comes from the combination of three MN disk potentials \citep{Smith2015}. This technique matches an exponential disk within 2\% out to 10 scale radii.  We adopt the fits provided by \citet{Smith2015} and sum three MN disks together to model the exponential stellar disk with our chosen scale height and scale length.

\subsubsection{Gas Disk}
\label{sssec:gas-disk}
The gaseous disk is modeled as an exponential by implementing the same triple MN disk technique discussed above \citep{Smith2015}.   The gas disk masses at high redshift are determined using the observational results of \citet{Popping2015} who provide gas fractions, $f_{g} = M_{\rm gas}/(M_{\rm gas} + M_{\star})$, for galaxies as a function of stellar mass.   Specifically, we use their median values for the cold gas fraction as a function of stellar mass and redshift to fit a 2-D regression. We then use this fit along with the stellar disk mass and redshift to set the cold gas mass. The scale lengths of the gas disk are fixed to be the same constant ratio with the stellar disk given at $z=0$.

\subsubsection{Stellar Bulge}
\label{sssec:bulge}
The bulge is modeled as a \citet{Hernquist90} potential where the scale length is a constant multiple of the stellar scale length as determined by both components' scale lengths at $z=0$. The bulge mass evolves to maintain the same ratio of bulge mass to stellar mass present at $z=0$.

\section{Results}
\label{sec:results}
Figure \ref{fig:density} shows example visualizations of the dark matter distribution for a typical halo in our suite simulated without (left) and with (right) the galaxy potential.  This is the halo identified as ``Kentucky" and ``Kentucky {\tt Disk}" in Table \ref{tab:sim-info}.   The top panels are 500 kpc boxes, and correspond approximately the virial volume of this halo (with $\rvir \simeq 270$ kpc). The lower panel is zoomed in to a region 100 kpc across. Qualitatively, our results are very similar to those of \citet{GarrisonKimmel2017}. The presence of a central galaxy eliminates a majority of the substructure in the innermost region ($\rm \leq 50~kpc$) but has only a minimal effect at large radius. The notable enhancement of dark matter the very center of the {\tt Disk} run is due to baryonic contraction. This effect is also apparent in full hydrodynamic simulations at this mass scale \citep{SGK2017}.

\subsection{Velocity Functions}
\label{ssec:vdist}
Figure \ref{fig:vfunc} shows the velocity functions for subhaloes in the DMO (black) and disk simulations (magenta).  Shown are cumulative $\rm \vmax$ distributions (left columns) and $\rm \vpeak$ distributions (right column) of all the resolved subhaloes within 300 kpc, 100 kpc, and 50 kpc in top, middle, and bottom rows, respectively. The bands bound the minimum and maximum values for each velocity bin and the thick lines represent the medians. The inclusion of a central galaxy potential (magenta) to the DMO runs affects subhaloes of all masses roughly uniformly and has a greater impact on the total number of subhaloes in regions closer to the disk.  Within $300$ kpc, the {\tt Disk} runs have $\sim 70\%$ the number of subhaloes seen in the DMO runs, roughly independent of velocity.  At $100$ kpc, the offset is close to a factor of $\sim 2$.  Within $50$ kpc, the difference is close to a factor of $\sim 3$.

One important feature seen in Figure \ref{fig:vfunc} is the roll-off in the $\vpeak$ functions at small velocity.  This is both a sign and measure of incompleteness.  Incompleteness in $\vpeak$ gets worse at smaller radius (where stripping is more important) as might be expected.   Within $100$ kpc (middle right) we show signs of incompleteness below $\vpeak \simeq 5~\kms$.   
Within $R=50$ kpc (bottom right) we appear complete for $\vpeak > 6 ~ \kms$.  Note that we show no major signs of completeness issues down to $\vmax = 4.5~\kms$ for all radii we have explored.  In Appendix \ref{append:supp} we present a resolution test using re-simulations of a DMO and \texttt{Disk} run with 64 times worse mass resolution.  Scaling from $\vmax = 4.5 ~\kms$ to the lower resolution simulation, we would expect convergence down to $\vmax \simeq 15 ~\kms$ (following the mass trend for subhalos, $M \propto \vmax^{3.45}$).  We indeed find agreement with the higher resolution simulation at $\vmax = 15 ~\kms$ in both the DMO and \texttt{Disk} resimulations.

\begin{figure}
	\hspace{-15pt} 
	\includegraphics[width=1.1\columnwidth]{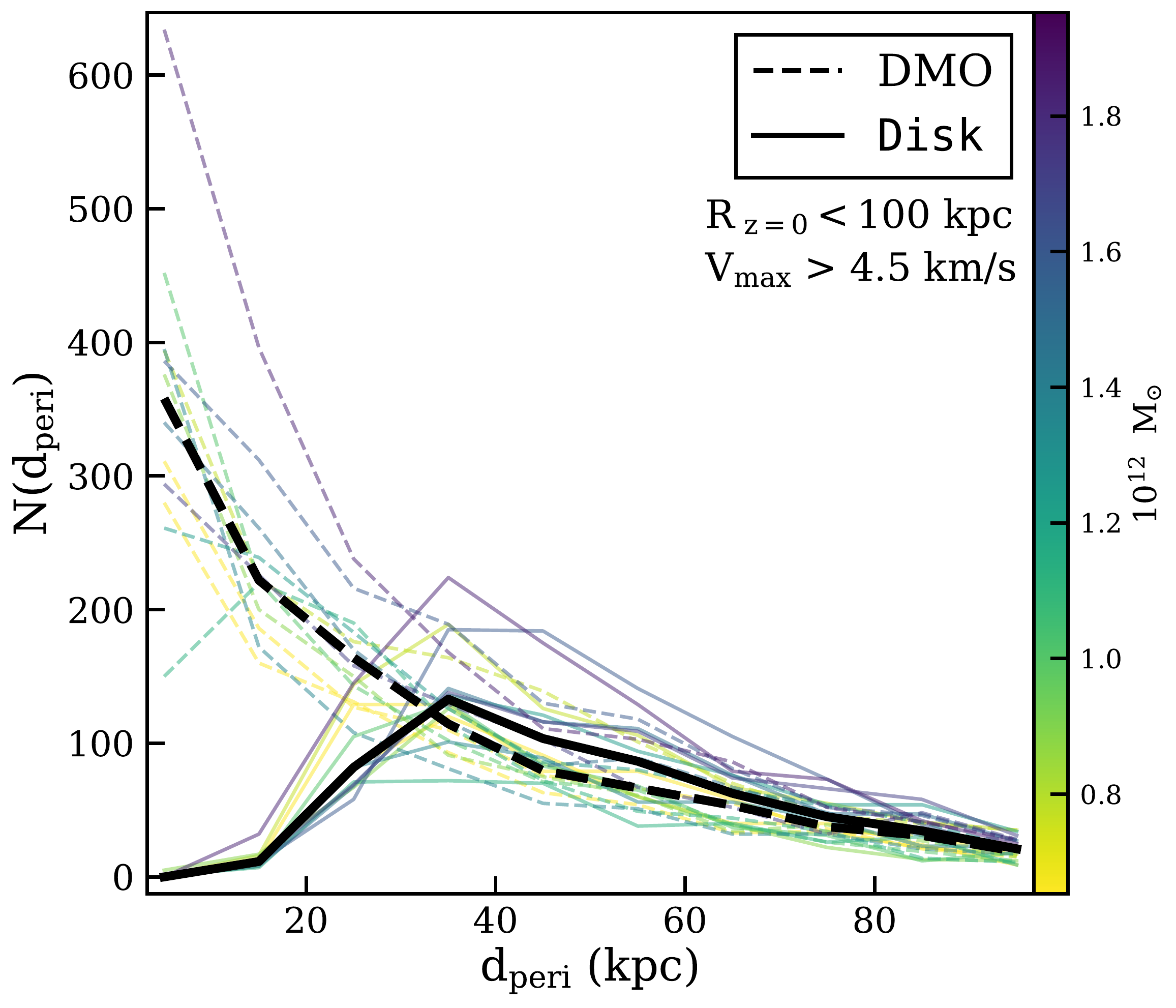}
    \caption{Distribution of the pericentric distances for all surviving subhaloes with $\vmax > 4.5 ~\kms$ within a present-day radius of $R=100$ kpc. The dashed and solid lines represent the subhalo distributions for individual host haloes in DMO and {\tt Disk} runs, respectively. The lines are colored according to host halo mass as indicated by the color bar on the right.  We do this to provide a way to help match haloes from one run to the next, not because there is any apparent trend with halo mass.  The thick lines show median relations for their respective simulation type. Note that the {\tt Disk} runs preferentially deplete subhaloes that have pericentres smaller than $\sim 20-30$ kpc. While the DMO simulations have pericentre distributions that spike towards zero, subhaloes in the {\tt Disk} runs have pericenter distributions that peak at $\sim 35$ kpc.}
    \label{fig:dperi}
\end{figure}

\subsection{Radial Distributions}
\label{ssec:rdist}
As seen in Figure \ref{fig:vfunc}, the difference between the DMO and {\tt Disk} runs increase with decreasing distance from the halo center.  This point is emphasized in Figure \ref{fig:rdist}, which shows cumulative radial profiles at fixed $\vmax$ (left) and $\vpeak$ (right) cuts in both DMO and {\tt Disk} runs.

\begin{figure*}
    \includegraphics[width=\columnwidth]{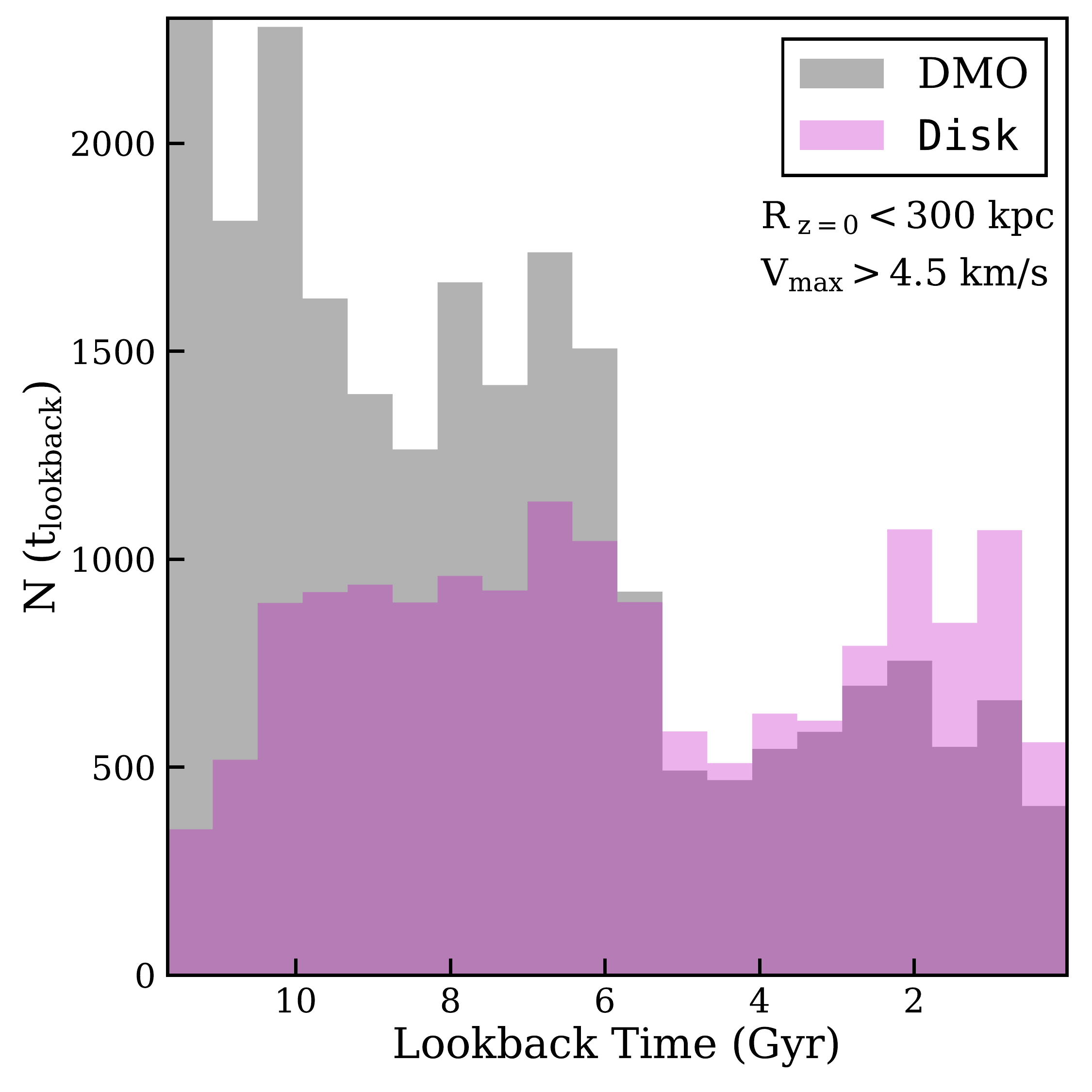}
    \includegraphics[width=\columnwidth]{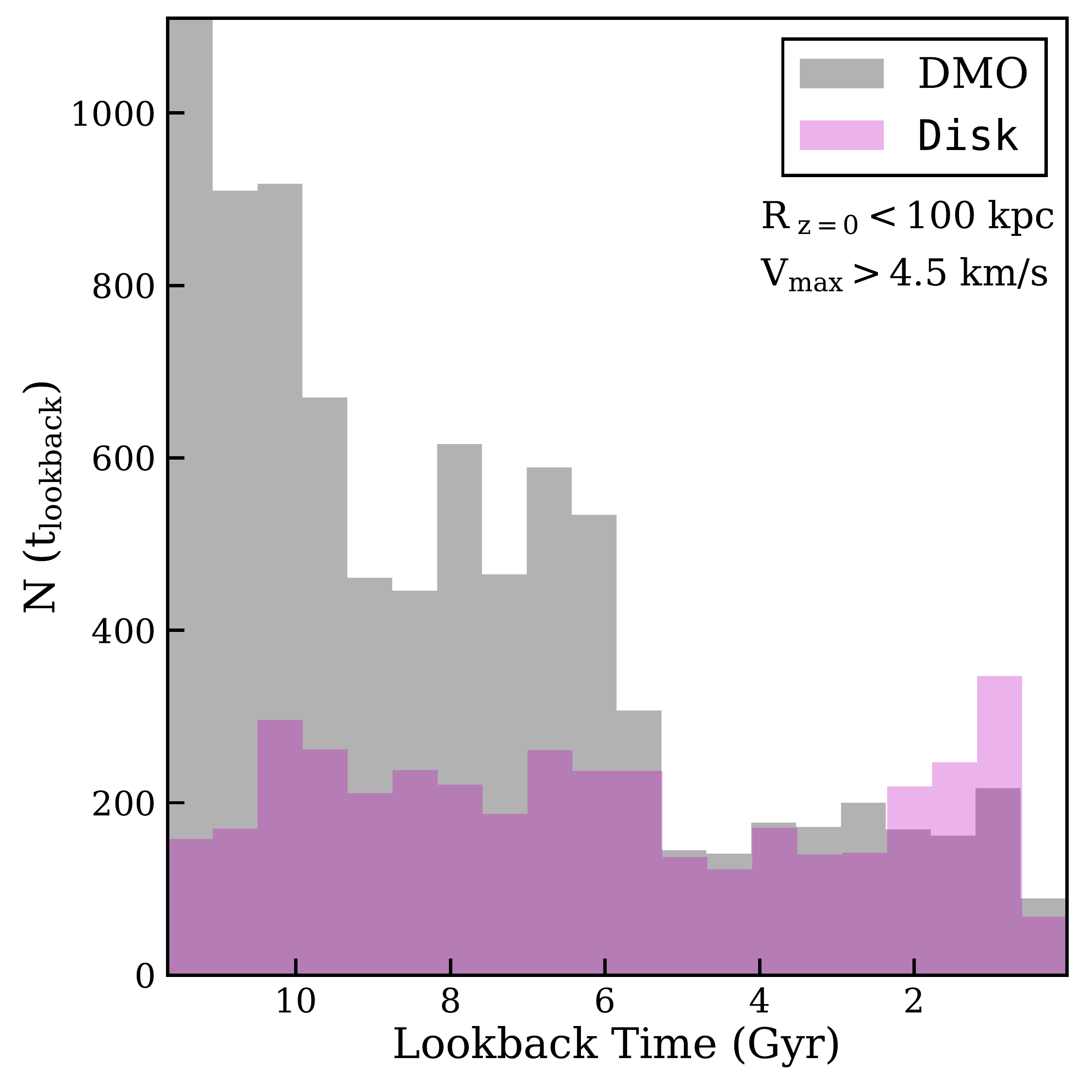}
    \caption{Distributions of infall times when subhaloes first crossed into the host virial radius.  Shown are distributions for all surviving subhaloes with $\vmax > 4.5 ~\kms$ within $R=300$ kpc (left) and $R=100$ kpc (right) stacked together for all of the DMO (gray) and {\tt Disk} (magenta) simulations in our suite.  The {\tt Disk} runs are clearly depleted in subhaloes that fell in more than $\sim 6$ Gyr ago compared to the DMO runs, and especially so in the $R<100$ kpc sample.}
    \label{fig:infall}
\end{figure*}

The left panel of Figure \ref{fig:rdist} shows the cumulative radial count of subhaloes with $\vmax > 4.5 ~\kms$ for each of our 12 DMO (black dash) and {\tt Disk} (magenta) hosts.  The thick black lines show medians for each of the distributions.  Note that while the difference in overall count is only $\sim 30\%$ out at $300$ kpc, the offset between DMO and {\tt Disk} grows to more than an order of magnitude at small radius, and is typically a factor of $\sim 20$ at $R=25$ kpc.  The majority of the {\tt Disk} runs have no identifiable substructure within $20$ kpc.  None of the {\tt Disk} simulations have even a single subhalo within $10$ kpc.  As can be seen in Table \ref{tab:sim-info}, the systematic depletion of central subhaloes occurs in every host, including the most massive halo (Hound Dog), where the ratio of galaxy mass to halo mass is the smallest. 

The right panel of Figure \ref{fig:rdist} tells a similar story. Here we have chosen a fairly large cut in $\vpeak > 10 ~\kms$. This scale is similar to, though somewhat smaller than, the natural scale where galaxy formation might naively be suppressed by an ionizing background \citep[e.g.][]{Okamoto08}.   The majority of the {\tt Disk} runs have nothing with $\vpeak > 10 ~\kms$ within $\sim 30$ kpc. As discussed in Section \ref{ssec:counts} and in \citet{Graus18b}, the fact that we already know of $5$ Milky Way satellites within 30 kpc of the Galactic center (and that we are not complete to ultra-faint galaxies over the full sky) suggests that we may need to populate haloes well smaller than this 'natural' scale of galaxy formation in order to explain the satellite galaxy population.

\subsection{Pericenter Distributions}
\label{ssec:peri}
At first glance, it is potentially surprising that the existence of a galaxy potential confined to the central regions of a halo can have such a dramatic effect on subhalo counts at distances out to $\sim 100$ kpc.  As first discussed by \citet{GarrisonKimmel2017}, the pericenter\footnote{Pericenters were obtained by interpolating the subhalo positions between snapshots and storing the minimum separation between the host and the subhalo as the pericenter. The time between interpolated snapshots is 14-16 Myr.} distribution of subhaloes provides some insight into this question.

Figure \ref{fig:dperi} shows the pericenter distributions of all subhaloes found within $R=100$ kpc at $z=0$ in both the DMO (dashed) and {\tt Disk} runs (solid).  There is a unique (thin) line for each halo, color coded by the halo virial mass (color bar).  The thick black lines are medians.  While the two distributions are similar for large pericenter differences ($R \gtrsim 40$ kpc) the differences are dramatic at $R \lesssim 20$ kpc.  Subhaloes in the DMO simulations exist on quite radial orbits, with $d_{\rm peri}$ distributions that spike towards $d_{\rm peri} = 0$.  Surviving subhaloes haloes in the {\tt Disk} runs, on the other hand, have distributions that peak at $d_{\rm peri} \sim 35$ kpc and have a sharp decline towards $d_{\rm peri} = 0$.   It is clear that subhaloes that get close to the galaxy potentials are getting destroyed.  

Figure \ref{fig:dperi} also shows that the differential effect of the disk potential on a given halo varies dramatically based on the underlying orbital distribution of its subhaloes.  DMO haloes that have the largest spike in low pericenters will have the largest overall shift in subhalo counts once disk potentials are included.  We find that for subhaloes that exist within 300 kpc but have never passed within 20 kpc, the difference in the radial and orbital distributions between the DMO and {\tt Disk} runs is negligible.

\subsection{Infall Times}
\label{ssec:infall}
The subhaloes that are present in the DMO runs but absent in the {\tt Disk} runs are biased not only in their orbital properties (Figure \ref{fig:dperi}) but in the time they have spent orbiting within their host haloes.  Figure \ref{fig:infall} shows the infall time distributions for subhaloes with $\vmax > 4.5 ~\kms$ for all of the DMO simulations (gray) and for the {\tt Disk} re-runs (magenta).  The left panel shows infall times for subhaloes that exist within $300$ kpc of their host halo centers at $z=0$.  The right panel shows infall times for subhaloes within $100$ kpc.  Times are plotted as lookback ages, with zero corresponding to the present day.

Both panels of Figure \ref{fig:infall} clearly demonstrate that subhaloes with early infall times are preferentially depleted in the {\tt Disk} runs.  The differences are particularly significant for infall times greater than $6$ Gyr ago: the early-infall tails are considerably depressed in the {\tt Disk}.
Interestingly, the shifts in median lookback times to infall are modest as we go from DMO to {\tt Disk}: $7.6$ Gyr to $6.1$ Gyr in the 300 kpc panel and $8$ Gyr to $6$ Gyr in the 100 kpc panel.   Also, the {\tt Disk} simulations show a slight {\it enhancement} of late-time accretions ($\sim 1-2$ Gyr).  This may be related to the halo contraction that occurs as the galaxy grows at late times (see concentration comparison in Table \ref{tab:sim-info}).  It is possible that some subhalos enter the viral volume faster than they would in the DMO equivalent because of this effect.  More analysis will be needed to test this hypothesis because the halo virial mass itself shows no such enhancement at late times.

\subsection{Time Evolution of Substructure Counts}
\label{ssec:streams}

Substructure in dark matter haloes is set by a competition between the accretion rate of small haloes and the mass loss rate from dynamical effects over time \citep[e.g.][]{ZB03}.  A central galaxy potential increases the destruction rate, which depletes subhalo populations compared to DMO simulations.

One question is whether and to what extent differences in subhalo counts seen between DMO and {\tt Disk} runs persists at earlier times.  This may have important observational implications for substructure probes that are sensitive properties at early times.  Cold stellar streams, for example, may have existed for multiple orbital times ($t_{\rm orb} \sim 500$ Myr).  If the substructure population was significantly higher in the past then this could manifest itself in observables today.

\begin{figure}
	\hspace{-15pt}
	\includegraphics[width=1.1\columnwidth]{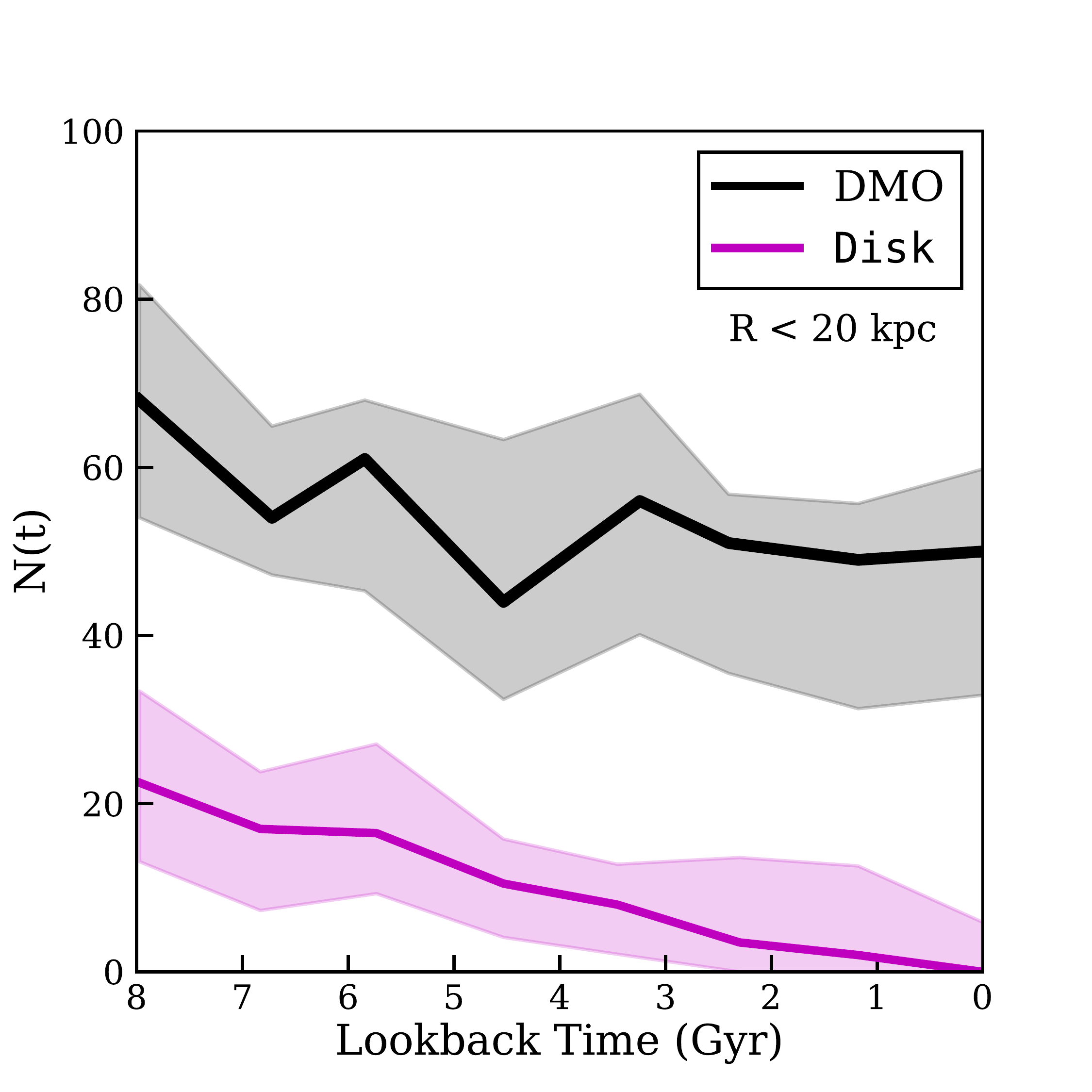}
    \caption{Number of subhaloes with $\vmax > 4.5 ~\kms$ that exist within 20 kpc (physical) of the host halo centre as a function of lookback time. The gray and magenta distributions represent the distributions for the DMO and {\tt Disk} runs, respectively.  The solid lines show medians and the bands cover the full spread of the data.  The difference between the two classes of runs persists back to $8$ Gyr.  }
    \label{fig:streams}
\end{figure}

Figure \ref{fig:streams} explores this question by showing the count of $\vmax > 4.5 ~\kms$ subhaloes within a physical radius of $20$ kpc of each halo center as a function of lookback time.  The bands show the full distributions over all simulations, with gray corresponding to DMO and magenta corresponding to the {\tt Disk} runs.  Solid lines are medians.  We see that the overall offset between DMO and {\tt Disk} runs persists to lookback times of $8$ Gyr, but that for times prior to $\sim 4$ Gyr ago, the subhalo counts in the {\tt Disk} runs begin to approach the DMO counts.  In the median, the difference is `only' a factor of $\sim 3$ eight billion years ago, compared to more than a factor of $\sim 30$ suppression at late times.

Overall, it appears that the expected suppression is quite significant in its implications for cold stellar stream heating.  The median count of subhaloes in the {\tt Disk} runs remains near zero over the past $\sim 2$ Gyr (compared to $\sim 50$ subhaloes in the DMO runs).  This timescale is $>3$ orbital times for a cold stream like Pal-5 at $R=20$ kpc and $V_{\rm orb} \sim 200 ~\kms$.  The median subhalo count in the {\tt Disk} runs remains less than ten to lookback times of $4$ Gyr.  Cold streams that have persisted for more than $4$ Gyr or extend out to $\sim 50~\kpc$ from the Galaxy may be required in order to provide robust probes of substructure, though a full exploration of this question will require work well beyond that presented in this introductory paper.  We hope that the public release of our subhalo catalogs will facilitate efforts of this kind.

\section{Implications}
\label{sec:imp}

\begin{figure}
	\hspace{-15pt} 
    \includegraphics[width=1.05\columnwidth]{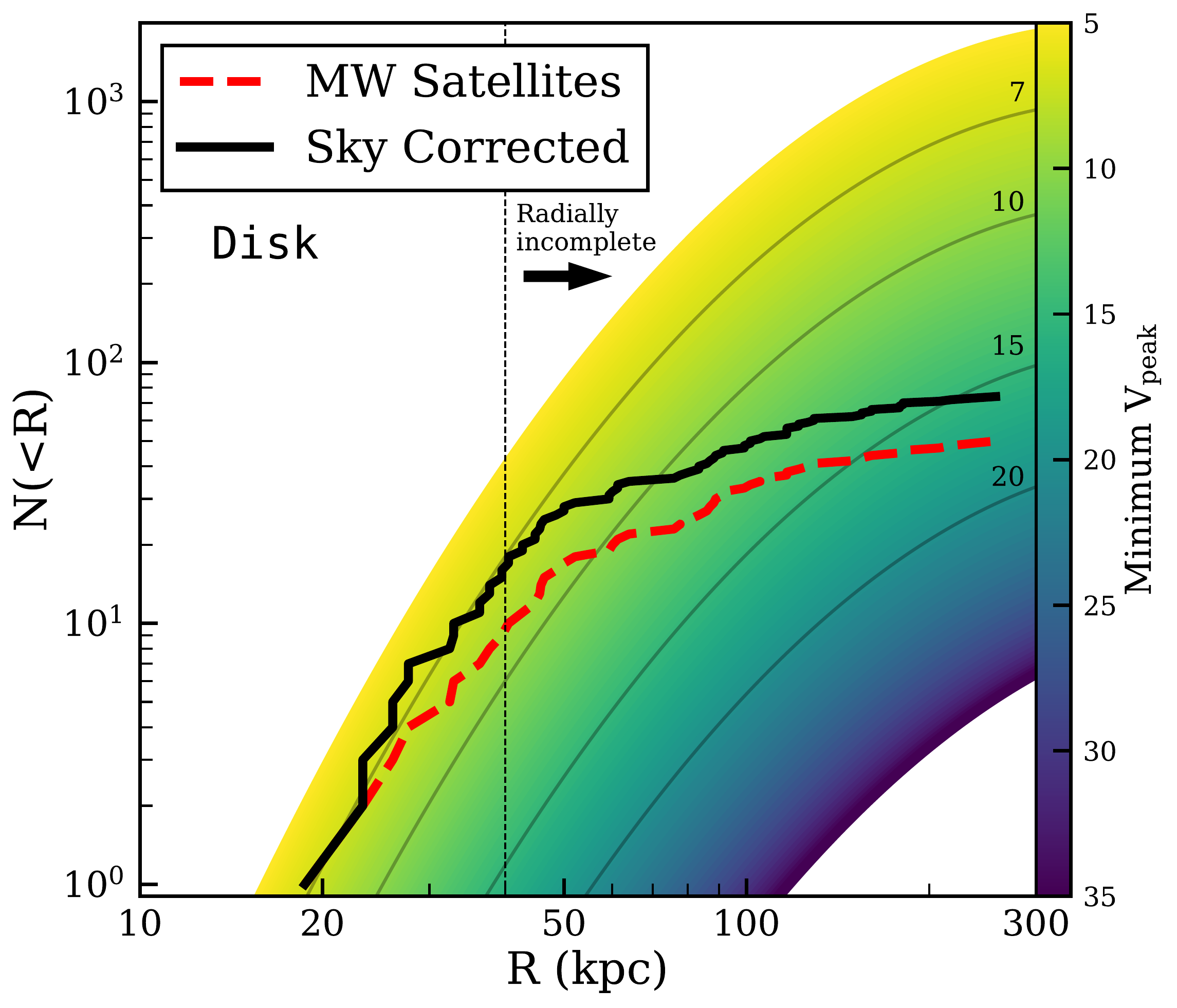}
    \caption{Median cumulative radial counts of subhalos for all of the {\tt Disk} runs color coded by $V_{\rm peak}$ threshold.  The faint gray lines mark $V_{\rm peak}$ thresholds larger than 7, 10, 15, and 20 $\kms$, respectively. Thick lines represent the Milky Way satellite data uncorrected (red dashed) and corrected for sky coverage (solid black). The vertical dotted line at 40 kpc represents an estimate of the radial completeness limit for $L \simeq 1000$ L$_{\odot}$ ultra-faint dwarfs. Observed counts to the right of this line should be treated as lower limits, as the true counts may be much higher than those shown given the lack of a deep, full sky survey. If our host halos are representative of the Milky Way, then we must populate all subhaloes with $\vpeak \gtrsim 7 ~\kms$ in order to account for the data.  This extrapolates to an implied total of $\sim 1000$ ultra-faint satellites within 300 kpc.}
    \label{fig:rad_means_disk}
\end{figure}

\begin{figure}
	\hspace{-15pt}
	\includegraphics[width=1.05\columnwidth]{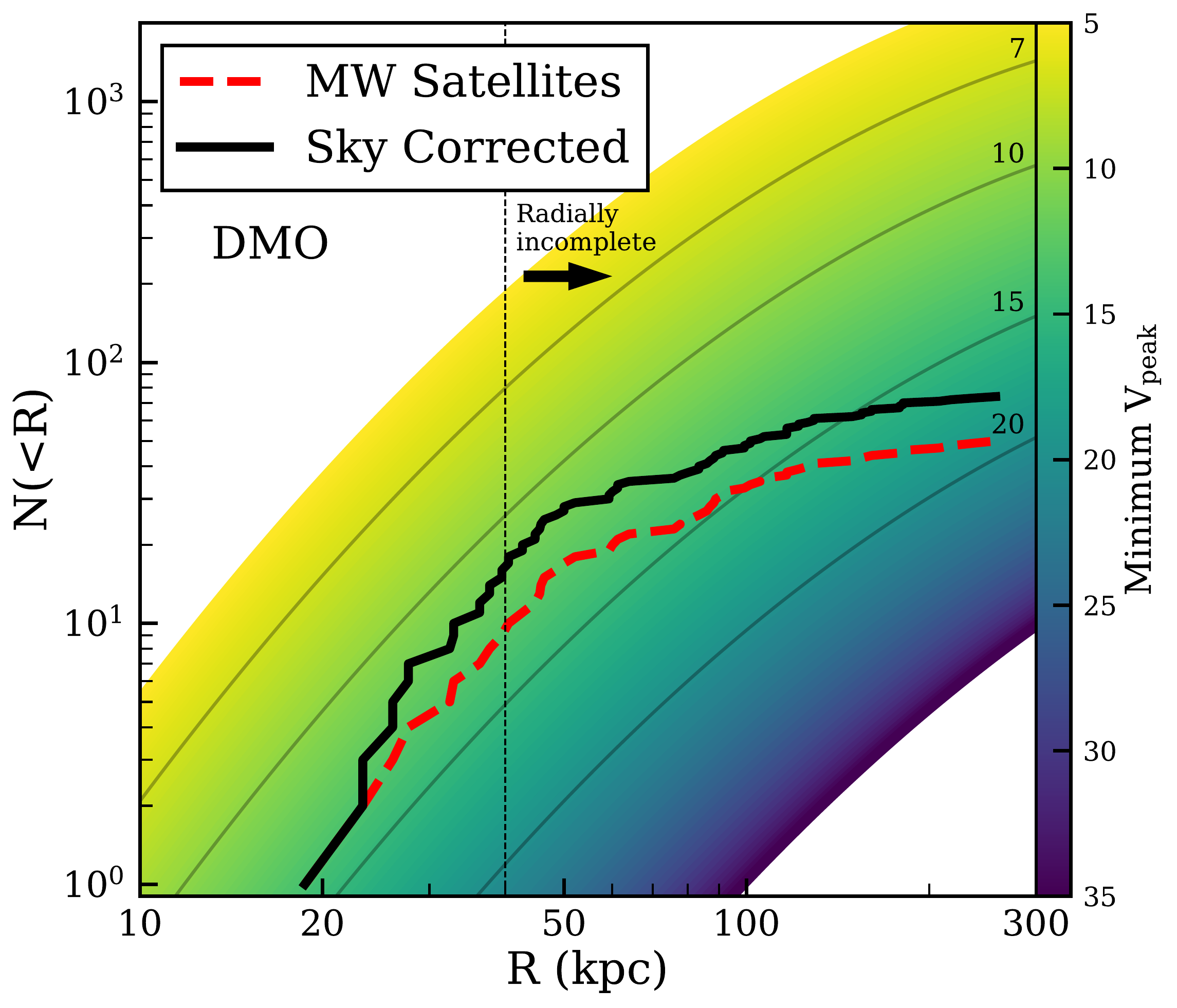}
    \caption{Median radial profiles for different $\vpeak$ cuts for all of the DMO runs. Compare to the {\tt Disk} simulations shown in Figure \ref{fig:rad_means_disk}. The faint gray lines represent the DMO data for fixed $\vpeak$ thresholds of 7, 10, 15, and 20 $\kms$, respectively. The thick lines represent the Milky Way satellite data uncorrected (red dashed) and corrected for sky coverage (solid black).  Haloes with $\vpeak \gtrsim 12 ~\kms$ are required to match the inner data in the median of our DMO runs, and this extrapolates to an implied total of $\sim 200$ ultra-faint satellites within 300 kpc.}
    \label{fig:rad_means_dmo}
\end{figure}

\subsection{What haloes host ultra-faint galaxies?}
\label{ssec:counts}

As alluded to in Section \ref{ssec:rdist}, the absence of substructure within the vicinity of the central galaxy in the {\tt Disk} runs may have important implications for our understanding of the mapping between galaxy haloes and stellar mass.  In particular, the relatively large number of galaxies that are known to exist within $\sim 50$ kpc of the Galactic center provides important information about the lowest mass dark matter haloes that are capable of forming stars \citep{Jethwa18}.

The majority of efforts to understand how the ionizing background suppresses galaxy formation have found that most dark matter haloes with $\vpeak < 20$ $\kms$ are devoid of stars  \citep[e.g.,][]{Thoul96,Okamoto08,Ocvirk16,Fitts17}. 
A second scale of relevance for low-mass galaxy formation is the atomic hydrogen cooling limit at $10^4$ K, which corresponds to a $V_{\rm peak} \simeq 16\,\kms$ halo. Systems smaller than this would require molecular cooling to form stars.  Taken together, one might expect that most ultra-faint satellite galaxies of the Milky Way should reside within subhaloes that fell in with peak circular velocities in the range $16-20\,\kms$.

Compare this basic expectation to the information summarized in Figure 
\ref{fig:rad_means_disk}.  Here we plot the {\it median} cumulative radial count of subhaloes with $\vpeak$ values larger than a given threshold as derived from the full sample of {\tt Disk} runs.   The color bar on the right indicates the $\vpeak$ threshold and the solid lines track characteristic $\vpeak$ values (7, 10, 15, and 20 $\kms$) as labeled. A similar figure that utilizes data from the DMO simulations is provided in Figure \ref{fig:rad_means_dmo}.

\begin{figure*}
	\includegraphics[width=\columnwidth]{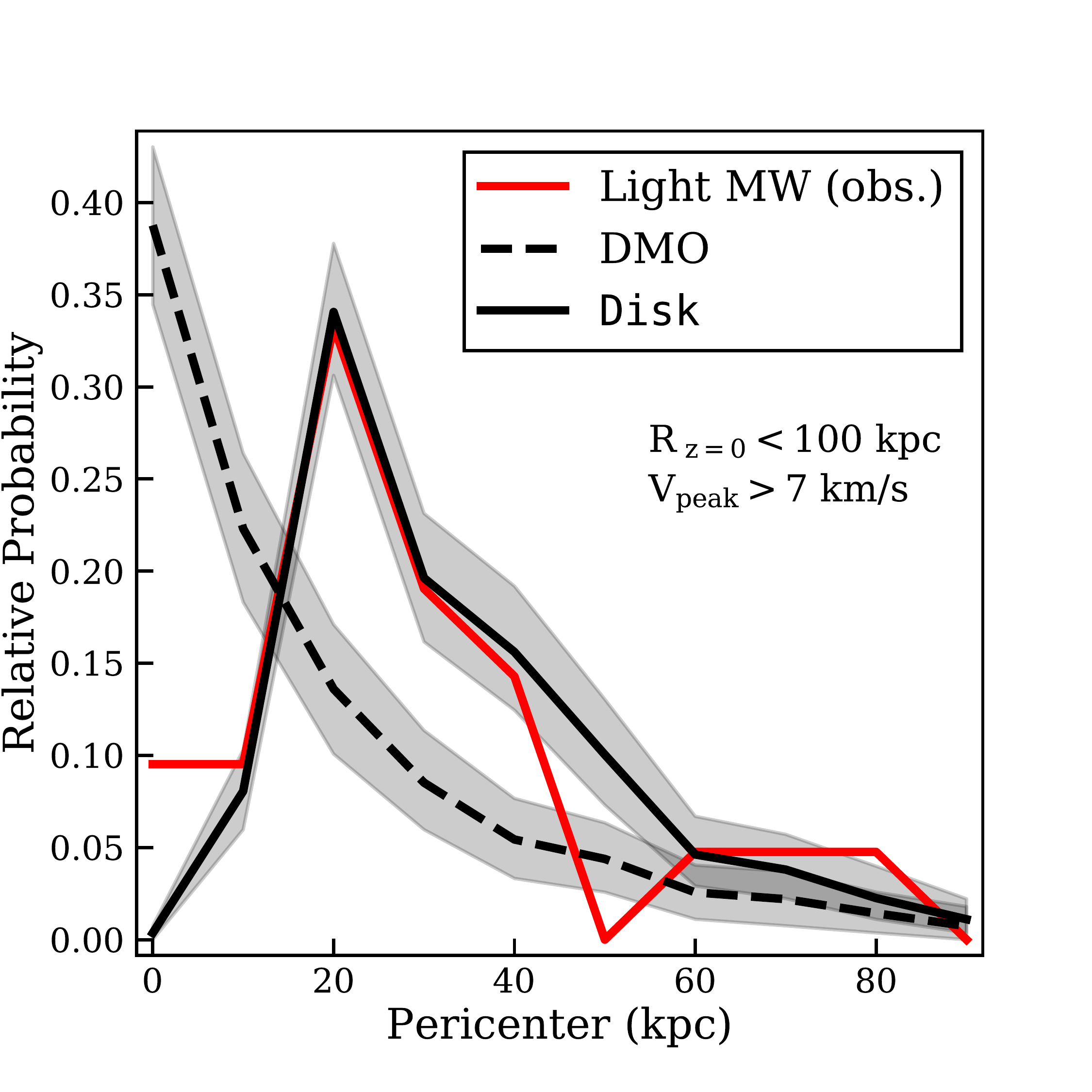}
    \includegraphics[width=\columnwidth]{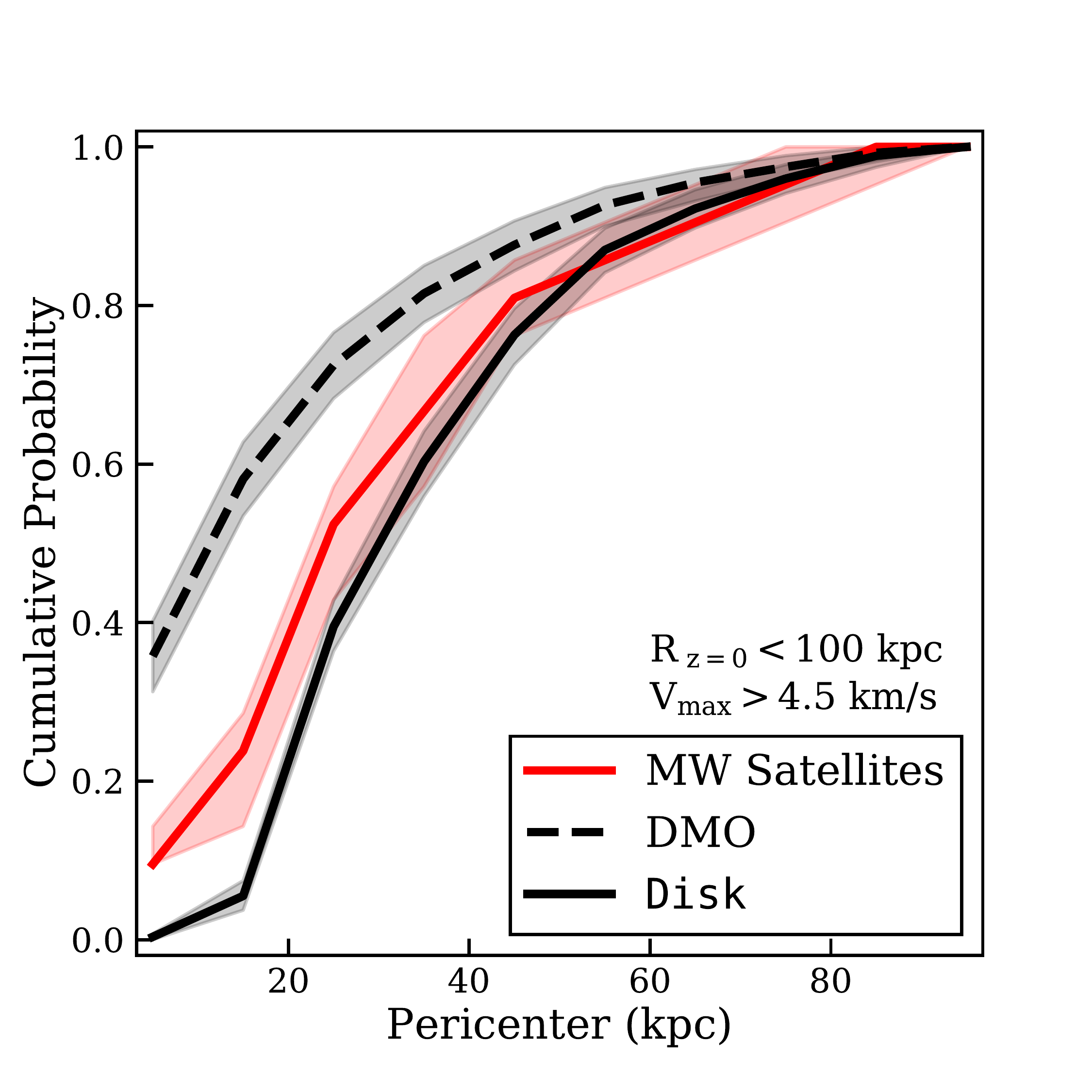}
    \caption{Comparison of the pericenter distributions for subhaloes (black) with $\vmax > 4.5~\kms$ to those of the 21 MW satellites (red) within 100 kpc of the Galactic center as derived by \citet{Fritz18a}. In order to account for radial completeness bias in the data, the subhaloes have been resampled to have the same $z=0$ radial distribution as the data. Each host halo is weighted equally.  Left: Differential DMO (dashed) and {\tt Disk} (solid) median pericentre distributions with the MW satellite pericentre distribution shown in red. The gray bands represent the 95\% confidence interval obtained from sampling the subhaloes pericentre distributions. Right: Cumulative pericentre distributions. The band represents the 95\% confidence interval obtained from sampling the MW satellite pericentre values given their respective errors.  Unlike the real MW satellites, the DMO subhaloes peak towards small pericentre.  The {\tt Disk} runs produce a subhalo pericentre distribution that is closer to the distribution seen in the data.}
    \label{fig:dperimw}
\end{figure*}

The dashed red line shows the census~\footnote{The \citet{Fritz18a} compilation does not include the LMC and SMC.  We also exclude their presence here to be conservative, as massive subhalos of this kind are rare in MW-size hosts and we are focusing primarily on implications for ultra-faint satellites.} of known MW satellites galaxies as compiled by \citet{Fritz18a}. The thick black line in Figure \ref{fig:rad_means_disk} applies a sky-coverage correction to derive a conservative estimate of the radial count of satellite galaxies. This correction assumes that 50\% of the sky has been covered by digital sky surveys to the depth necessary to discover ultra-faint galaxies and adds a second galaxy for every MW dwarf known that has an absolute magnitude fainter  than -6.  Importantly, even in the region of the sky that has been covered by digital sky surveys like SDSS and DES, our census of the faintest ultra-faint galaxies ($L \lesssim 10^3$ L$_\odot$) is not complete at radii larger than $\sim 40$ kpc \citep{Walsh2009}.  We draw attention to this fact with the vertical dotted line.

If our simulation suite is indicative of the Milky Way, we must associate the galaxies within $30$ kpc with subhaloes that had maximum circular velocities at infall greater than just $7 ~\kms$ \citep[corresponding to a peak infall mass of $M_{\rm peak} \simeq 3 \times 10^7 \Msun$,][]{ELVIS}. This is not only well below the canonical photo-ionization suppression threshold ($\sim 20 ~\kms$), it is smaller than the atomic cooling limit ($\sim 16 ~\kms$).  The virial temperature of the required $\sim 7 ~\kms$ haloes is $2000$ K, which likely would need efficient molecular cooling for star formation to proceed. If we perform the same exercise host-by host, the minimum $\vpeak$ required to explain the galaxy counts within 40 kpc varies some.  Nine of our 12 \texttt{Disk} runs require $\vpeak = 6.5-7.5$ km s$^{-1}$ to explain the counts within 40 kpc.  The other three require $\vpeak = 8.1$, $9.2$, and $9.3$ km s$^{-1}$, respectively.    We find no trend between the minimum $\vpeak$ required to explain the known counts and host halo mass.  In a companion paper by \citet{Graus18b}  we explore the implications of this basic finding and provide a statistical comparison based on each of our {\tt Disk} runs individually.  

In addition to changing our basic picture of low-mass galaxy formation, the need to populate tiny $\vpeak = 7 ~\kms$ haloes with galaxies means that there should be a very large number of ultra-faint galaxies within the virial radius of the Milky Way.  By tracking the $7 ~\kms$ line out to 300 kpc in Figure \ref{fig:rad_means_disk}, we see that it reaches $\sim 1000$ such objects.  If they are there in such numbers, future surveys like LSST should find them.  There would of course be many more outside of the virial radius.  In the field, the number density of these tiny haloes is $\sim 100$ Mpc$^{-3}$ \citep[e.g.][]{BBK17}. This means that there may be $100,000$  ultra-faint galaxies for every $L_*$ galaxy in the universe.

Figure \ref{fig:rad_means_dmo} is analogous to Figure \ref{fig:rad_means_disk} except now we compare the cumulative count of known MW galaxies to predictions for the DMO runs. There are many more haloes at small radii than in the {\tt Disk} runs and this means that to account for the number of galaxies seen within $\sim 40$ kpc we can populate more massive systems: $\vpeak \gtrsim 12 ~\kms$ halos.  If this were the case, we would expect only $\sim 200$ ultra-faint galaxies to exist within $300$ kpc of the Milky Way, which is in line with older expectations for satellite completeness limits based on DMO simulations \citep{Tollerud2008}.  It is interesting that the slopes of the predicted cumulative counts in Figure \ref{fig:rad_means_disk} are more similar to the observed radial profile of satellites within $\sim 50$ kpc than the profiles in the DMO runs shown in Figure \ref{fig:rad_means_dmo}.  This is perhaps an indication that by including the existence of the Galactic disk,  we are approaching a more accurate model of the Milky Way's satellite population.

\subsection{Satellite Pericenters}
\label{ssec:sat_peris}

As we discussed in reference to Figure \ref{fig:dperi}, subhalo pericenter distributions are dramatically different once the galaxy potential is included.  Here we take advantage of recent insights on satellite galaxy orbits made possible by \textit{Gaia} to determine which of these distributions is more in line with observations \citep{Gaia18b,Simon18,Erkal18,Pace2018,Fritz18a}. 
 
Figure \ref{fig:dperimw} presents a comparison of subhalo pericenters in the DMO (dashed) and {\tt Disk} runs (solid) to those of MW satellite galaxies. Shown in red are the differential (left) and cumulative (right) pericenter distributions of MW galaxies from \citet{Fritz18a} that have $z=0$ distances within 100 kpc of the Galactic center.  The \citet{Fritz18a} sample includes proper motions of 21 satellite galaxies within 100 kpc. Two MW potentials are used in \citet{Fritz18a} to derive the pericenter distances of each satellite. They are based on the \textit{MWPotential14} potential (see \citealt{}{Bovy 2015} for details) with a light and heavy DM halo with virial masses of $0.8 \times 10^{12} \Msun$ and $1.6 \times 10^{12} \Msun$, respectively. For clarity, we only include the results of the ``Light" MW potential, which is closer to the median halo mass of our sample ($1.1 \times 10^{12} \Msun$).  Results for the ``Heavy" MW potential  are very similar and can be found in Appendix \ref{append:supp}.

In order to fairly compare predictions to observations in this space, we must account for observational incompleteness.  Our current census of faint galaxies is radially biased within $100$ kpc, such that we are missing galaxies at large radii.  In order to make a fair comparison, we took all subhaloes with $\vmax > 4.5 ~\kms$ within a $z=0$ distance of 100 kpc of the center of each halo and then subsampled those populations 1000 times for each halo to create present-day radial distributions that match those of the satellites in \citet{Fritz18a}. We then ``stacked" these populations together to derive median pericenter distributions for subhaloes in each of the two classes of simulations (DMO and {\tt Disk}).  Note that each host halo is equally weighted. 

The left panel of Figure \ref{fig:dperimw} compares the median of the radially re-sampled distributions to the distribution of pericenters derived by \citet{Fritz18a}.  As foreshadowed in Figure \ref{fig:dperi}, the DMO subhaloes have a pericenter distribution that spikes towards small values, very unlike the distribution seen in the real data.  The {\tt Disk} runs, on the other hand, show a peak at $\sim 30$ kpc with rapid fall-off at smaller radii and a more gradually fall-off towards larger distances.  This shape is quite similar to that seen in the real data. Note that if we choose subhaloes with $\vpeak > 7 ~\kms$ instead, the distributions are almost indistinguishable (see Appendix \ref{append:supp}). It is interesting that the total lack of subhaloes with pericenters smaller than $20$ kpc is \textit{not} seen in the data.  The two galaxies in this inner bin are Segue 2 and Tucana III; these systems may very well be in the process of disruption. 

In the right panel of Figure \ref{fig:dperimw} we present the same data cumulatively and also explore how uncertainties in the derived orbits affect the comparison.  Specifically, we used the quoted errors given by \citet{Fritz18a} on each galaxy and drew from a Gaussian to generate $10,000$ realizations for each system.  The median of the resultant distribution is given by the thick red line with 95\% confidence intervals shown by the shaded band.  The DMO distribution is well above the 95\% region everywhere within $50$ kpc. The median of {\tt Disk} runs remain within the spread for all but the inner most region.  

From the above comparison, we conclude that the DMO runs produce a pericenter distribution for satellite subhaloes that is quite far from what is observed for Milky Way satellite galaxies.  The {\tt Disk} runs are much closer to what is observed and therefore appear to provide a more realistic comparison set.  The clear next step in this comparison is to re-derive the implied pericenter distributions for each host halo's mass and to directly compare predictions in full phase space to those observed.  While such an analysis is beyond the scope of this introductory paper, future work in this direction is warranted.  Understanding how host halo-to-halo scatter, ongoing satellite disruption,  and specifics of halo finding affect these interpretations will also be important.

\section{Discussion and Conclusions}
\label{sec:disc}

In this paper, we have introduced Phat ELVIS suite of 12 high-resolution simulations of Milky Way mass dark matter haloes that are each run with ({\tt Disk}) and without (DMO) a Milky Way disk galaxy potential.  As summarized in Table 2, the host halo masses in our suite span $M_{\rm vir} = 0.7-2 \times 10^{12} \Msun$, which encompasses most recent estimates for the virial mass of the Milky Way \citep{BlandHawthorn2016}.  The galaxy potential at $z=0$ is the same for each {\tt Disk} run and is summarized in Table 1. As demonstrated in Figure \ref{fig:vfunc}, our resolution allows us to have convergence in identifying subhaloes down to a  maximum circular velocity of $\vmax = 4.5~\kms$ ($M \simeq 5 \times 10^6 \Msun$) and with peak (infall) circular velocities $\vpeak \simeq 6 ~\kms$ ($M_{\rm peak} \simeq 1.8 \times 10^7 \Msun$). 
The main effect of the Milky Way potential on subhalo populations is that subhaloes with pericenters smaller than $\sim 20 ~\kpc$ are depleted in the {\tt Disk} runs (see Figure \ref{fig:dperi}).     

\subsection{Impact of the Disk on substructure populations}
The most striking difference between the {\tt Disk} and DMO subhaloes is in their abundances at radii smaller than $\sim 50 ~\kpc$ at $z=0$.  This difference can be seen visually in Figure \ref{fig:density} and quantitatively in 
Figure \ref{fig:rdist}.  Table \ref{tab:sim-info} lists counts as a function of various radial choices and shows that the ratio of subhalo counts between the {\tt Disk} to DMO runs at $z=0$ is typically $\sim 1/10$ at $R < 25 ~\kpc$, $\sim 1/3$ at $R< 50 ~\kpc$, and $\sim 1/2$ at $R< 100 ~\kpc$.  Note that these ratios are fairly constant independent of the host halo virial radius (or concentration).  To zeroth order, the depletion radius appears to be set by the disk potential (which is the same for all runs), not host halo properties.  The most important predictor for relative depletion seems to be the variable pericenter distributions in the DMO runs: simulations that have subhaloes with an over-abundance of percienters smaller than $\sim 20$ kpc will experience more relative depletion once the galaxy potential is included. 

Another difference between the surviving subhalo populations in the DMO and {\tt Disk} runs is in the distribution of infall times (see Figure \ref{fig:infall}).  If the galaxy potential is included, the majority of subhaloes that fell in more than $\sim 8$ Gyr ago and survived in the DMO runs become destroyed in the {\tt Disk} runs. This may have important implications for models of environmental galaxy quenching when applied to the Milky Way \citep{RW18,Fillingham2015,Wetzel2015,Wheeler2014} and may also potentially change the expected mapping between orbital energies and infall time expected for Milky Way satellites \citep{Rocha2012}.

\subsection{Numerical Convergence}
Before moving on to summarize some potential observational implications of our results, it is worth discussing numerical completeness.   Figure \ref{fig:vfunc} provides evidence that the mass functions are converged for subhaloes with infall masses down to $M_{\rm peak} \simeq 1.8 \times 10^7 \Msun$ ($\rm N_p \sim 600$ particles).  This level of completeness is typical of that quoted for simulations of this kind \citep[e.g.][]{Aquarius,ELVIS}. In Appendix \ref{append:supp} we present a resolution test using a re-simulation of one of our halos with 64 times worse mass resolution, and show that we are indeed converged to subhalos that are 64 times more massive than we have estimated in the high-resolution runs.  We also show using this low-resolution comparison that there is not a significant difference in convergence between the DMO and \texttt{Disk} runs.  This suggests that the offset between our \texttt{Disk} and DMO subhalo distributions is a real, physical effect.

While we have shown convergence, it is important to remind ourselves that convergence to an answer does not necessarily imply convergence to the correct answer.    Such a concern is raised by \citet{vdB18a} and \citet{vdB18b}, who have performed numerical experiments showing that many more particles may be required for robust tracking of subhalo disruption. For example, \citet{vdB18a} find that orbits passing within 10-20\% of the virial radius of a host (30-60 kpc for our haloes) may require $\rm N_p > 10^6$ particles for an accurate treatment.  For our simulations, this would correspond to subhaloes with mass $\sim 3 \times 10^9 \Msun$ or $\vmax \sim 30 ~\kms$.  As can be seen in Figure \ref{fig:vfunc}, even at this mass scale our simulations still show significant differences between the DMO and {\tt Disk} runs at small radius, and at roughly the same ratios reported for the lower-mass regime.  More work will be required to understand the origin of the puzzling differences between our naive understanding of convergence and the detailed work by \citet{vdB18b} to thoroughly understand subhalo mass loss.

\subsection{Observable consequences}
Modulo the above concerns about potential completeness issues, the simulation suite presented here has produced a number of results with potentially interesting implications for interpreting observations.  
\begin{itemize}
\item The majority of the {\tt Disk} simulations have no subhaloes larger than $V_{\rm max} = 4.5$ km s$^{-1}$ within $20$ kpc (Figure \ref{fig:rdist}) and the overall count of subhaloes within this radius remains depressed compared to the DMO runs for several billion years in the past (Figure \ref{fig:streams}).  This suggests that local stream-heating signals from dark substructure may be quite rare, even in cold dark matter models without suppressed small-scale power spectra.

\item  The pericenter distributions of Milky Way satellites derived from \textit{Gaia} data are remarkably similar to the pericenter distributions of subhaloes in the {\tt Disk} runs, while the DMO runs drastically over-predict galaxies with pericenters smaller than 20 kpc (Figure \ref{fig:dperimw}).  This suggests that the Galaxy potential must be considered in any attempt to understand the dynamics and evolution of Milky Way satellites, especially those that exist within the inner $\sim 100$ kpc of the Milky Way.

\item  As shown in Figure \ref{fig:rad_means_disk}, the depletion of inner substructure in the {\tt Disk} runs presents a tension with satellite galaxy counts that is in the opposite sense as that in the Missing Satellites Problem.  In order to account for all of the ultra-faint galaxies known within $40$ kpc of the Galaxy, we must populate haloes well below the atomic cooling limit ($\vpeak \simeq 7$ km s$^{-1}$ or $M \simeq 3 \times 10^{7} \Msun$ at infall). The precise value for the minumum $\vpeak$ varies from host to host, with 9 of our 12 \texttt{Disk} runs requiring $\vpeak = 6.5-7.5$ km s$^{-1}$ to explain the counts within 40 kpc.  The other three require $\vpeak = 8.1$, $9.2$, and $9.3$ km s$^{-1}$, respectively.  There is no apparent trend with host halo mass in the derived minimum values.  This issue is discussed in more detail in a companion paper by \citet{Graus18b}.

\item If tiny $\vpeak \simeq 7 ~\kms$ haloes do host ultra-faint galaxies, as implied by Figure \ref{fig:rad_means_disk}, this implies the existence of at least $\sim 1000$ satellite galaxies within 300 kpc of the Milky Way. The number density of such tiny haloes is $\sim 100$ Mpc$^{-3}$ \citep[e.g.][]{BBK17} in the field, suggesting that there may be $\sim 100,000$  ultra-faint galaxies for every $L_*$ galaxy in the universe.
\end{itemize}

The aim of this simulation suite is to provide a more accurate set of predictions for dark subhalo properties by including the inevitable existence of a central galaxy potential in calculations of their dynamical evolution.  We have focused here on a Milky Way galaxy analog in order to make direct connections to the well-studied population of Milky Way satellites.   A similar approach could be used to model satellite subhalo populations for a diverse set of galaxies.   

We have shown that the presence of the galaxy significantly changes our expectations for subhalo counts, orbits, and dynamical evolution and that this has a direct bearing on our interpretation of observed satellite galaxy properties as well as efforts to find dark subhalos.  Future work in this direction may prove vital in efforts to constrain the nature of dark matter and the physics of galaxy formation on the smallest scales.

\section*{Acknowledgements}

The authors would like to thank Sean Fillingham, Michael Cooper, Alex Drlica-Wagner, Denis Erkal, and Josh Simon for useful discussions.  TK and JSB were  supported by NSF AST-1518291, HST-AR-14282, and HST-AR-13888.  MBK acknowledges support from NSF grant AST-1517226 and CAREER grant AST-1752913 and from NASA grants NNX17AG29G and HST-AR-13888, HST-AR-13896, HST-AR-14282, HST-AR-14554, HST-AR-15006, HST-GO-12914, and HST-GO-14191 from the Space Telescope Science Institute, which is operated by AURA, Inc., under NASA contract NAS5-26555. Support for SGK was provided by NASA through the Einstein Postdoctoral Fellowship grant number PF5-160136 awarded by the Chandra X-ray Center, which is operated by the Smithsonian Astrophysical Observatory for NASA under contract NAS8-03060. MSP acknowledges that support for this work was provided by NASA through Hubble Fellowship grant \#HST-HF2-51379.001-A awarded by the Space Telescope Science Institute, which is operated by the Association of Universities for Research in Astronomy, Inc., for NASA, under contract NAS5-26555. AGS was supported by an AGEP-GRS supplement to NSF grant AST-1009973. Numerical business was taken care of in a flash using computational resources of the Texas Advanced Computing Center (TACC; http://www.tacc.utexas.edu), the NASA Advanced Supercomputing (NAS) Division and the NASA Center for Climate Simulation (NCCS), and the Extreme Science and Engineering Discovery Environment (XSEDE), which is supported by National Science Foundation grant number OCI-1053575. This work also made use of Astropy \footnote{\url{https://www.astropy.org}}, a community-developed core Python package for Astronomy \citep{Astropy, Astropy18}, matplotlib \citep{Matplotlib}, numpy \citep{Numpy}, scipy \citep{scipy}, ipython \citep{ipython}, pandas \citep{pandas}, Mayavi \citep{mayavi}, and the NASA Astrophysics Data System.



\appendix

\section{Supplementary Information}
\label{append:supp}

\begin{figure}
	\includegraphics[width=\columnwidth]{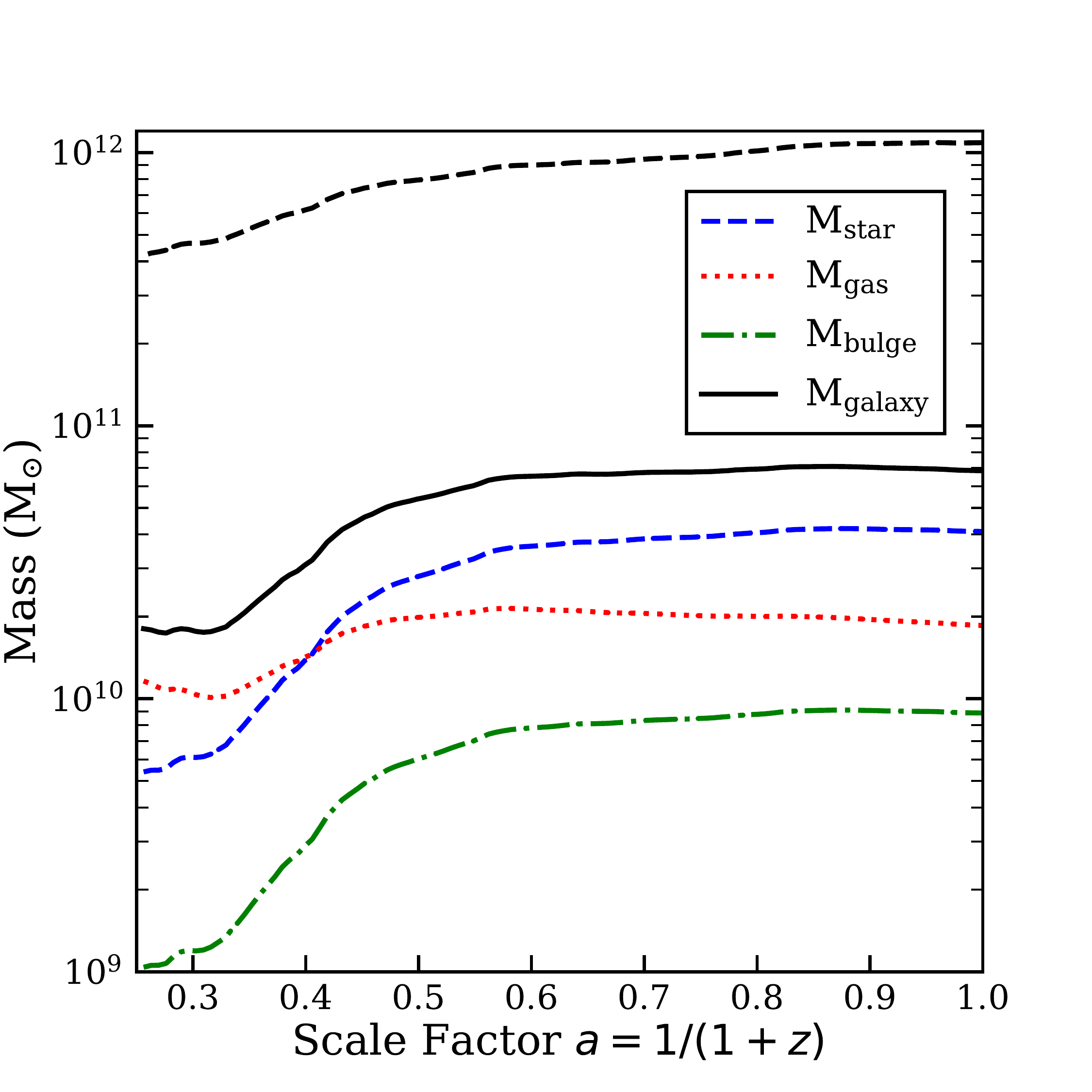}
    \caption{Mass growth of the galaxy for Kentucky with scale factor. The individual components' growths as well as the total galaxy mass growth. The final ($a=1$) position of all galaxy lines (not the dark matter halo virial mass) is fixed for all hosts as discussed in section \ref{ssec:potentials}. The dashed line near the top of the figure shows the halo's virial mass evolution.}
    \label{fig:growgal}
\end{figure}

Figure \ref{fig:growgal} provides an example growth history for Kentucky in order to illustrate how our galaxy components are evolved.  The dashed black line shows the main progenitor halo growth.  The solid black line shows the growth of the full galaxy mass.  The stellar disk (blue dashed), gas disk (red dotted), and bulge (green dash-dot) are forced to the values listed in Table \ref{tab:mw-values} at $z=0$.  The stellar mass (disk plus bulge) is set to track the host halo growth using abundance matching.    The gas disk masses at high redshift are determined using the observational results of \citet{Popping2015} who provide gas fractions for galaxies as a function of stellar mass.   

\begin{figure*}
	\includegraphics[width=\columnwidth]{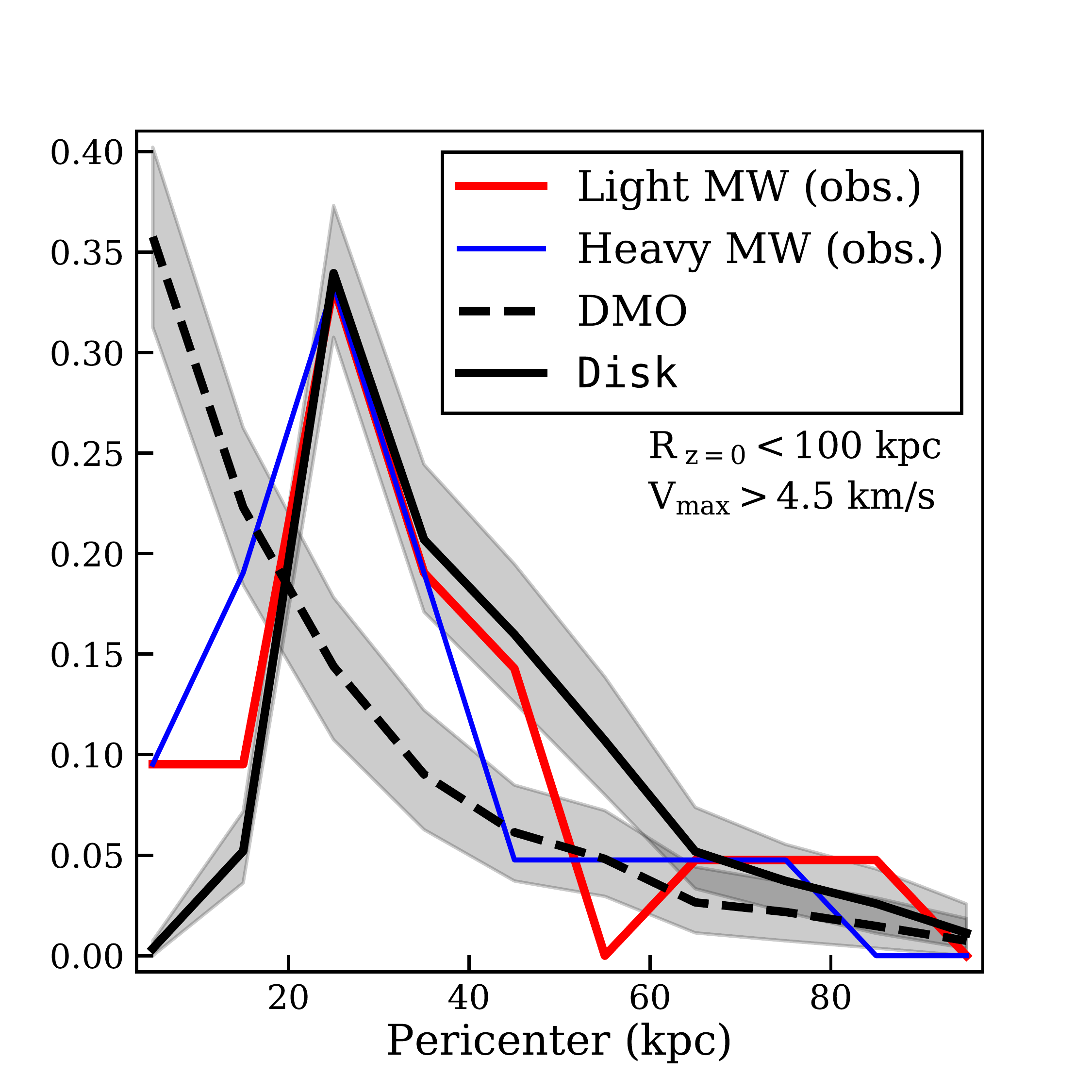}
    \includegraphics[width=\columnwidth]{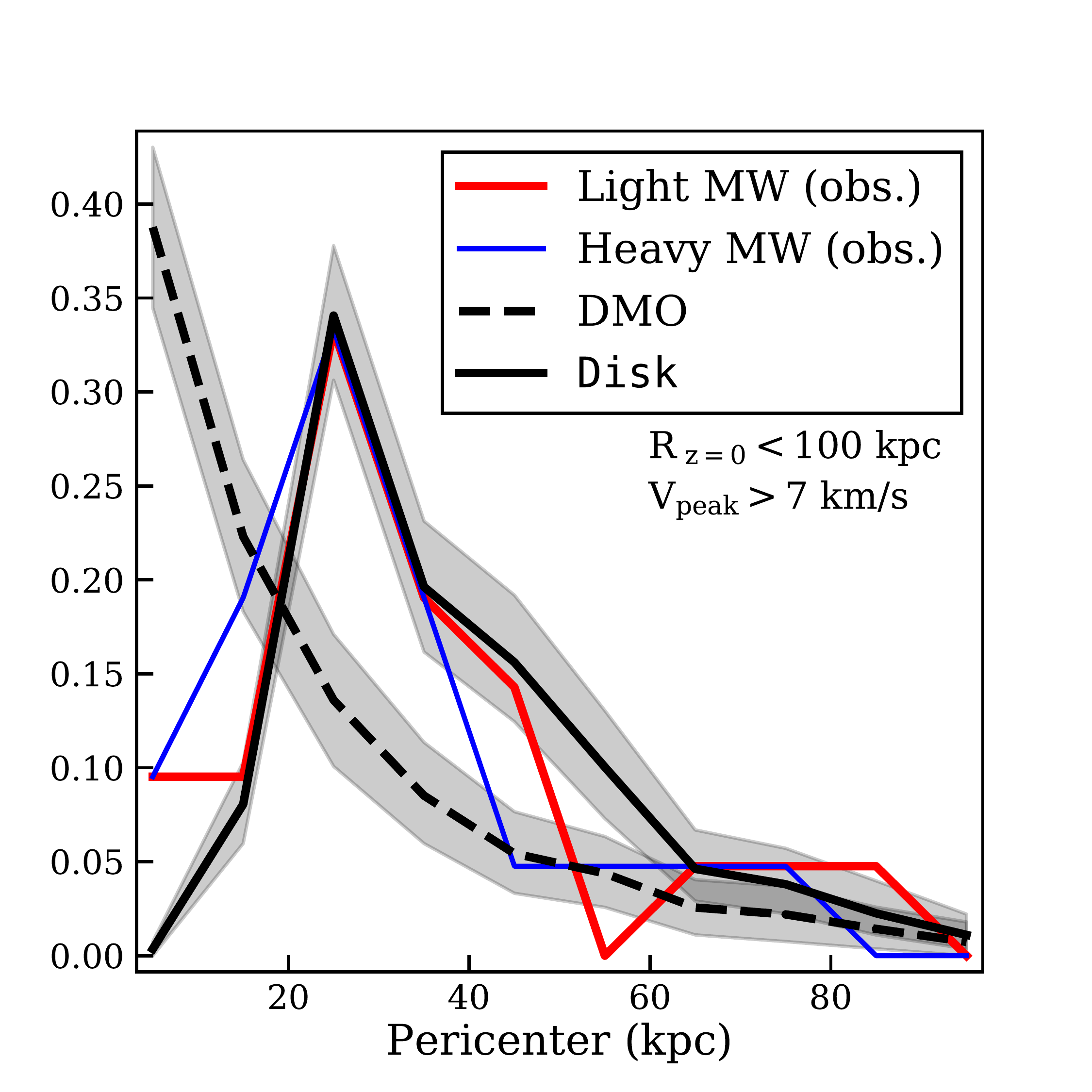}
    \caption{Differential pericentre distributions for subhaloes (black) with $\vmax > 4.5~\kms$ (left) and $\vpeak > 7~\kms$ (right) and MW satellites derived from both potentials used in \citet{Fritz18a}. The pericentres derived using the ``Light" MW potential are shown by the red line and is the same as figure \ref{fig:dperi} while those derived using the ``Heavy" MW potential are shown by the blue line. The gray bands represent the 95\% confidence interval for the subhaloes' distributions.}
    \label{fig:subperis}
\end{figure*}

Figure \ref{fig:subperis} is analogous to Figure \ref{fig:dperimw} in that it compares the pericenter distributions of subhaloes to those of Milky Way satellite galaxies presented \citet{Fritz18a}. Here we include the pericenters derived using both the ``Light" (red) and ``Heavy" (blue) MW potential in \citet{Fritz18a}.   We also show the subhalo distributions for a $\vmax$ cut ($ > 4.5 ~\kms$, left) and $\vpeak$ cut ($> 7 ~\kms$, right).  These different choices do not change the qualitative result that the observed satellite distributions are closer to the {\tt Disk} runs than the DMO runs. 

\begin{figure*}
    \centering
    \includegraphics[width=\columnwidth]{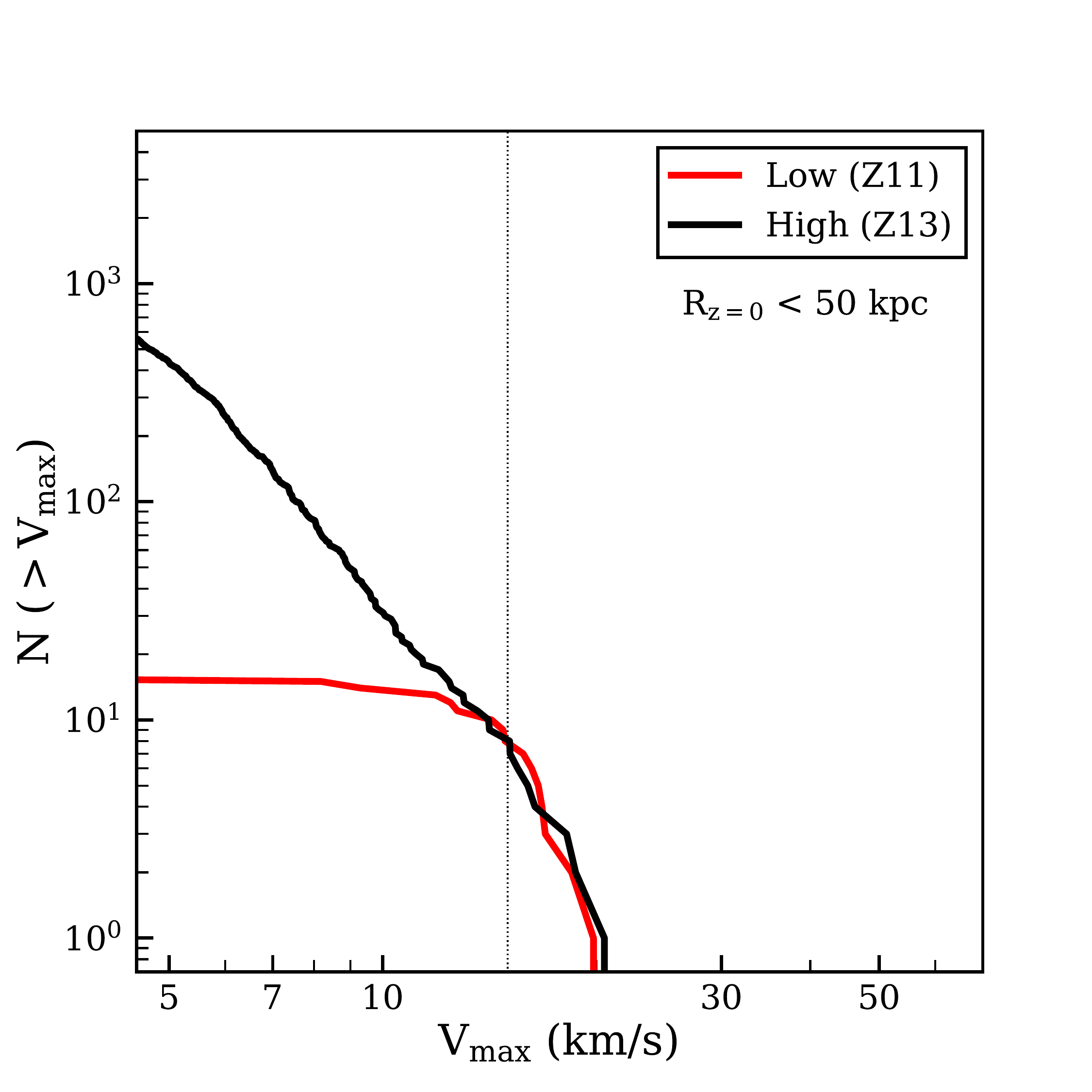}
    \includegraphics[width=\columnwidth]{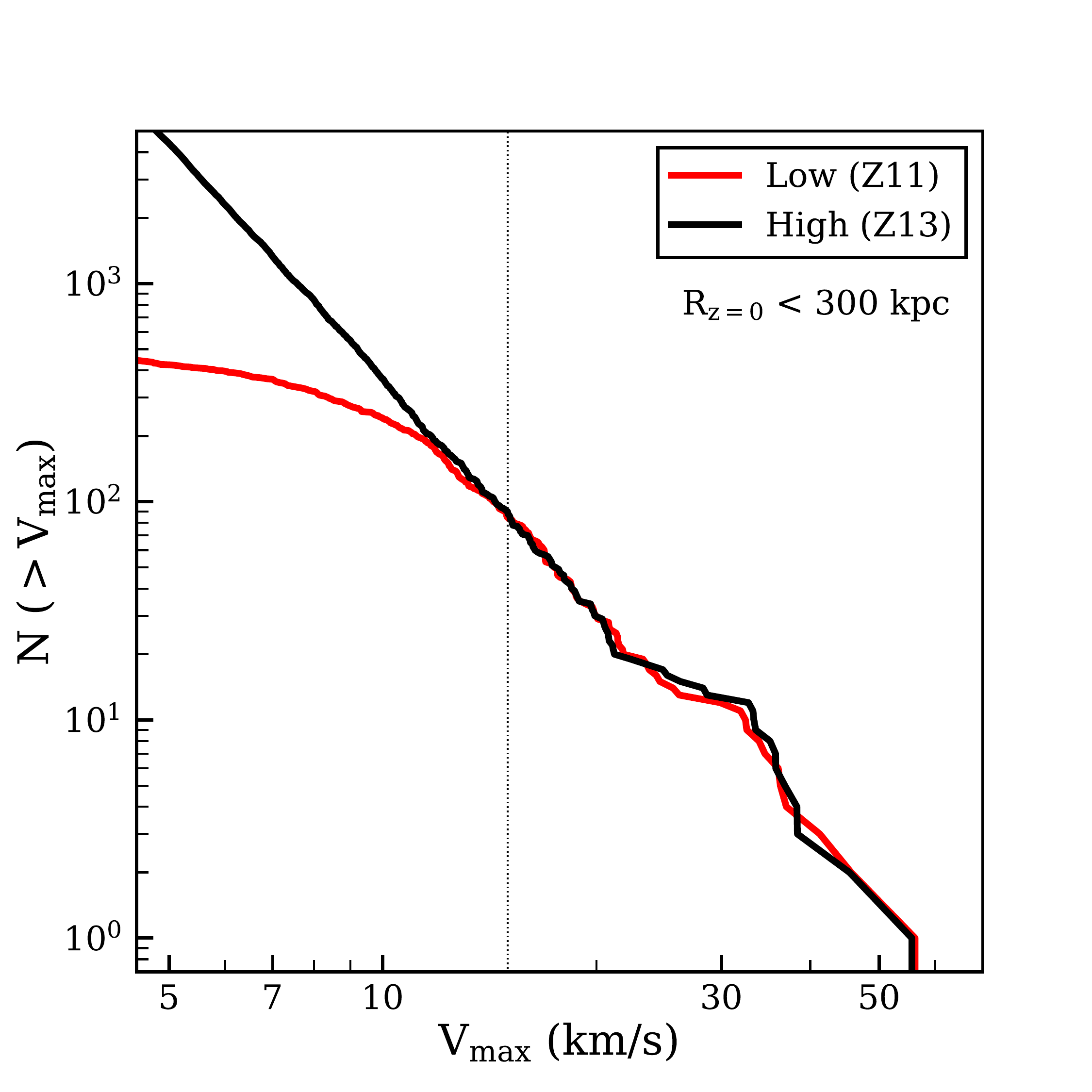}
    \caption{Comparison of the subhalo counts of the Hound Dog host halo from our fiducial (High, black) run to a lower resolution run (Low, red), with $64\times$ fewer particles. The left panel shows a cumulative count of all subhaloes above a given $\vmax$, within 50 kpc of the host centre. The right panel is similar but includes subhaloes out to 300 kpc. The vertical dotted line indicates where we would expect convergence ($\vmax = 15 ~\kms$) by scaling our adopted completeness threshold for the fiducial high-resolution runs ($\vmax = 4.5 ~\kms$). }
    \label{fig:converge}
\end{figure*}

\begin{figure*}
    \centering
    \includegraphics[width=\columnwidth]{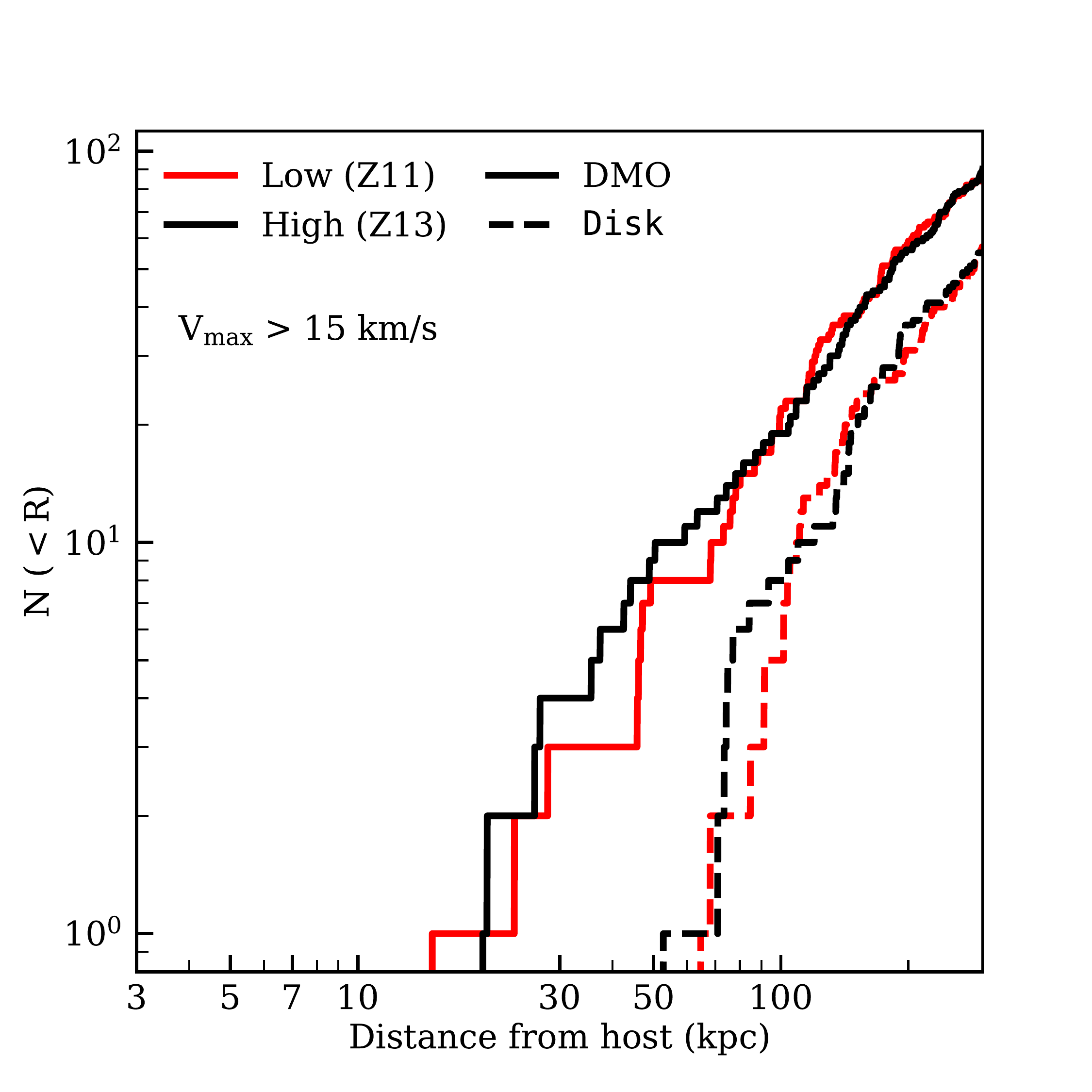}
    \caption{Radial radial distributions of subhaloes with $\vmax > 15 ~\kms$ for the Hound Dog host halo from the fiducial resolution (High, black) and a run with $64\times$ fewer particles (Low, red). Runs with Milky Way potentials \texttt{Disk} are shown as the dashed lines. The counts  agree to within expected Poisson variation for both types of runs at all radii.}
    \label{fig:conv_rad}
\end{figure*}

Figure \ref{fig:converge} illustrates the effects of numerical resolution on the $\vmax$ function for the `Hound Dog' host halo. The black line shows the results obtained from our fiducial resolution for all objects within 50 kpc (left panel) and 300 kpc (right panel). The red line shows the results obtained from the same halo rerun with $64\times$ fewer particles. The estimated completeness for our high resolution runs used in the main paper is $\vmax = 4.5 ~\kms$ (corresponding to subhalos with $\sim 170$ particles, see \ref{ssec:vdist}).  Using $M \propto \vmax^{3.45}$, we would expect the lower resolution comparison to be complete to $\vmax \simeq 15 ~\kms$ at fixed particle count.  We note that the two simulations do indeed begin to systematically differ only below $\vmax \simeq 15 ~\kms$, which is indicated by the vertical dotted line. Figure \ref{fig:conv_rad} shows the radial distributions of subhaloes with $\vmax > 15 ~\kms$ from our high-resolution and low-resolution runs both with (dashed) and without (solid) embedded galaxy potentials.  The two resolutions are consistent to within counting errors at all radii. 
Importantly, the DMO and \texttt{Disk} runs appear to be converged down to the same $\vmax$, which suggests that the differences we see with and without the galaxy potential are real, physical differences and not associated with spurious numerical effects.

\bibliographystyle{mnras}
\bibliography{elvis_disk}

\bsp	
\label{lastpage}
\end{document}